\documentclass[12pt]{article}
\pdfoutput=1

\RequirePackage[round]{natbib}
\usepackage{epsfig}
\usepackage{boxedminipage}
\usepackage{multirow}
\usepackage{amsmath}
\usepackage{eepic}
\usepackage{amssymb}
\usepackage{color}
\usepackage{url}
\usepackage{epic,ecltree}
\usepackage{longtable}
\usepackage{pgfpages}
\usepackage{lscape}
\usepackage{multirow}
\usepackage{makeidx} 
\usepackage{type1cm} 
\usepackage[bottom]{footmisc}
\usepackage{float}
\usepackage{subfig}
\RequirePackage[dvips,a4paper,textwidth=16.5cm,textheight=22.9cm]{geometry}

\usepackage{bm}

\usepackage{hyperref}
\definecolor{light-gray}{gray}{0.95}

\newcommand{\ed}{\end{document}}

\newcommand{\vy}{{\bm y}}
\newcommand{\vz}{{\bm z}}

\makeindex

\begin{document}
\title{Relabelling in Bayesian mixture models by pivotal units}
\author{Leonardo Egidi\thanks{Dipartimento di Scienze Statistiche, Universit\`a degli Studi di Padova, Italy, e-mail:egidi@stat.unipd.it} \and Roberta Pappad\`{a} \thanks{Dipartimento di Scienze Economiche, Aziendali, Matematiche e Statistiche `Bruno de Finetti', Universit\`{a} degli Studi di Trieste, Via Tigor 22, 34124 Trieste, Italy, e-mail: rpappada@units.it, francesco.pauli@deams.units.it, nicola.torelli@deams.units.it} \and Francesco Pauli\footnotemark[2] \and Nicola Torelli\footnotemark[2]}
\date{}

\maketitle






\vspace{1cm}

\begin{abstract} 
In this paper a simple procedure to deal with label switching when exploring complex posterior distributions by MCMC algorithms is proposed.
Although it cannot be generalized to any situation, it may be handy in many applications because of its simplicity and very low computational burden.
A possible area where it proves to be useful is when deriving a sample for the posterior distribution arising from finite mixture models, when no simple or rational ordering between the components is available. 
\end{abstract}

\section{Introduction}

Label switching is a well-known and fundamental problem in Bayesian estimation
of finite mixture models \citep{McPeel2000}.
The label switching problem arises when exploring complex posterior distributions by Markov Chain Monte Carlo (MCMC) algorithms because the likelihood of the model is invariant to the relabelling of mixture components.

Since there are as many maxima as there are permutations of the indices ($G!$), the likelihood has then multiple global maxima. 
This is a minor problem (if a problem at all) when we perform classical inference, since any maximum leads to a valid solution and inferential conclusions are the same regardless of which one is chosen.

On the contrary, invariance with respect to labels is a major problem when Bayesian inference is used. 
If the prior distribution is invariant with respect to the labelling as well as the likelihood, then the posterior distribution is multimodal.

To make inference on a parameter specific of a component of the mixture, a sample from the posterior that represent different modes would be inappropriate. An actual MCMC sample may or may not switch labels depending on the efficiency of the sampler. If the raw MCMC sampler randomly switches labels, then it is unsuitable for exploring the posterior distributions for component-related parameters.


A range of solutions has been proposed to perform inference in presence of label
switching \citep{fruhwirth2001markov, stephens2000dealing}, but for complex models most of the existing procedures are complex and computationally expensive.

A full solution entails obtaining valid samples for each parameter, and the methods in Section~\ref{sec:relabeling} are designed to relabel the raw Markov chains for this purpose. Simpler solutions are available if we do not need posterior samples for all the parameters.

In this paper a simple procedure based on the post MCMC relabelling of the chains to deal with label switching when exploring complex posterior distributions by MCMC algorithms is proposed.

 As pointed out in Section~\ref{sec:mcmc}, we can totally ignore the relabelling if the quantities of interest are label invariant. 
Besides the extreme case of label invariant quantities, we illustrate in Section~\ref{sec:clustering} and \ref{sec:probgroups} how to obtain a clustering and even a matrix of probabilities of units belonging to groups, using the raw MCMC sample without the need to fully relabel it.
In Section~\ref{sec:pivotal} we propose a method which performs a relabelling starting from a suitable clustering of the samples, with the aim of using an MCMC sample to infer on the characteristics of the components in terms of both probabilities of each unit being in each group and the group parameters. The performance of the algorithm is explored via the simulation study discussed in Section~\ref{sim:study} and a case study on real data is presented in Section~\ref{case:study}. Section~\ref{conc} concludes.

\newcommand{\vmu}{{\bm \mu}}
\newcommand{\vpi}{{\bm \pi}}
\newcommand{\ca}[1]{\ensuremath{[#1]_h}}
\newcommand{\uno}[1]{\ensuremath{\left|#1\right|}}
\newcommand{\data}{\ensuremath{\mathcal D}}

\section{The relabelling problem}
\label{sec:problem}
Prototypical models in which the labelling issue arises are mixture models, where, for a sample $\vy=(y_1,\ldots,y_n)$ we assume
\begin{equation*}
 (Y_i|Z_i=g) \thicksim f(y;\mu_g,\phi),  \label{eq:ydist} 
\end{equation*}
where the $Z_i$, $i=1,\ldots,n$, are i.i.d.\ random variables and
\begin{equation*}
 Z_i\in\{1,\ldots,G\},\;\; P(Z_i=g)=\pi_g. \label{eq:gdist}
\end{equation*}
The likelihood of the model is then
\begin{equation}
L(\vy;\vmu,\vpi,\phi) = \prod_{i=1}^n \sum_{g=1}^G \pi_g f(y_i;\mu_g,\phi),
\label{eq:lik}
\end{equation}
with $\vmu=(\mu_{1},\dots,\mu_{G})$ component-specific parameters and $\vpi=(\pi_{1},\dots,\pi_{G})$ mixture weights. Equation~\eqref{eq:lik} is invariant under a permutation of the indices of the groups, that is,  if
$(j_1,\ldots,j_G)$ is a permutation of $(1,\ldots,G)$ and $\vpi'=(\pi_{j_1},\ldots,\pi_{j_G}) $, $\vmu'=(\mu_{j_1},\ldots,\mu_{j_G}) $ are the corresponding permutations of $\vpi$ and $\vmu$, then
\begin{equation}
L(\vy;\vmu,\vpi,\phi) = L(\vy;\vmu',\vpi',\phi). \label{eq:invlik}
\end{equation}
As a consequence, the model is unidentified with respect to an arbitrary permutation of the labels.

When Bayesian inference for the model is performed, if the prior distribution $p_0(\vmu,\vpi,\phi)$ is invariant under a permutation of the indices, that is $p_0(\vmu,\vpi,\phi) = p_0(\vmu',\vpi',\phi)$, then so is the posterior 
\begin{equation}
p(\vmu,\vpi,\phi|\vy) \propto p_0(\vmu,\vpi,\phi)L(\vy;\vmu,\vpi,\phi),
\end{equation}
which is then multimodal with (at least) $G!$ modes. This implies that all simulated parameters should be switched to one among the $G!$ symmetric areas
of the posterior distribution, by applying suitable permutations of the labels to each MCMC draw.

\subsection{Relabelling and label switching in MCMC sampling}
\label{sec:mcmc}
In the following we assume that we obtained an MCMC  sample from the posterior distribution for model (\ref{eq:lik}) with a prior which is labelling invariant.
We denote as $\{\ca{\theta}:h=1,\ldots,H\}$ the sample for the parameter $\theta=(\vmu,\vpi,\phi)$.
We assume that also the $Z$ variable is MCMC sampled and denote as $\{\ca{Z}:h=1,\ldots,H\}$ the corresponding sample.

\begin{figure}
\begin{center}
\begin{tabular}{ccc}
{\it (a)} & {\it (b)} & {\it (c)} \\
\includegraphics[width=0.3\textwidth]{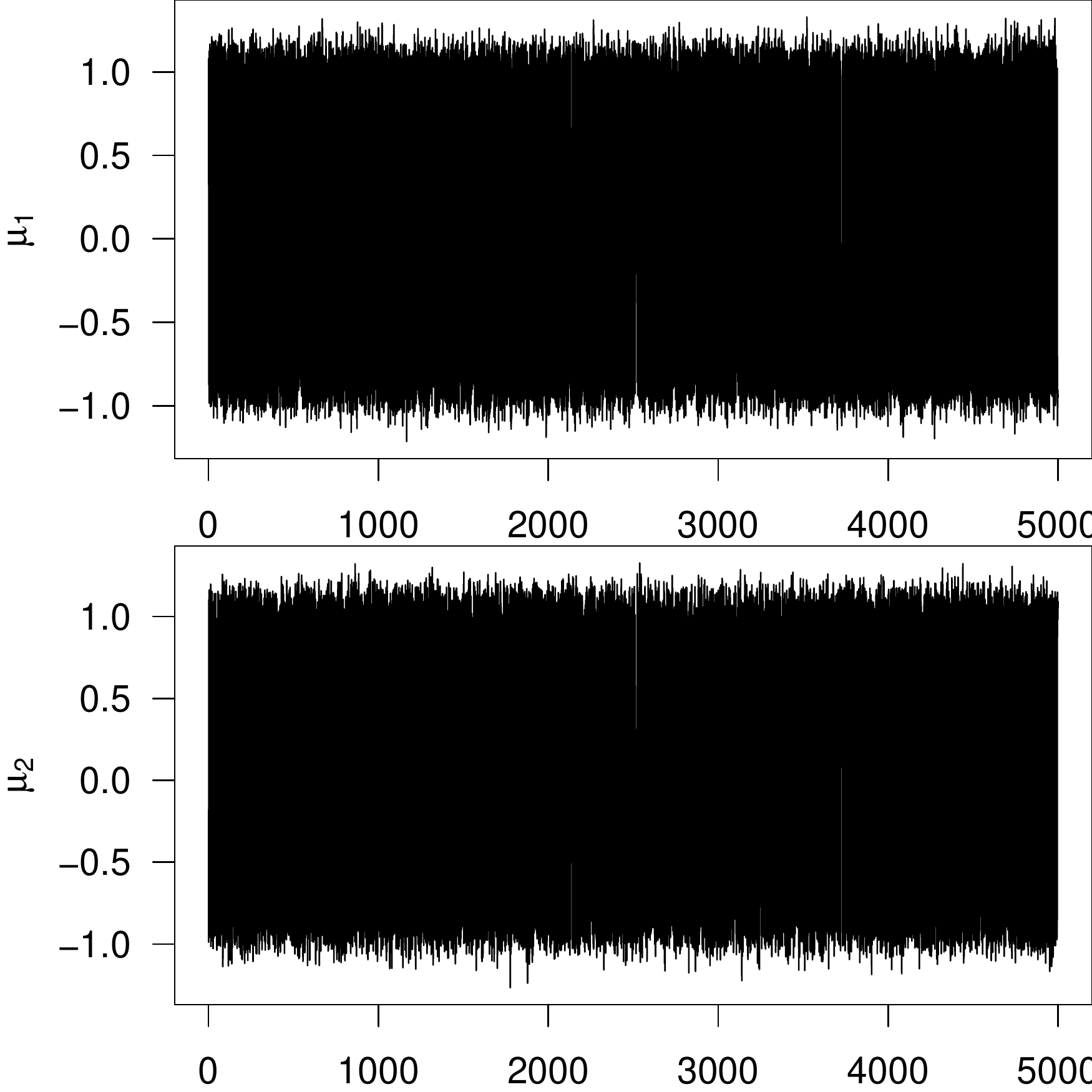}
&
\includegraphics[width=0.3\textwidth]{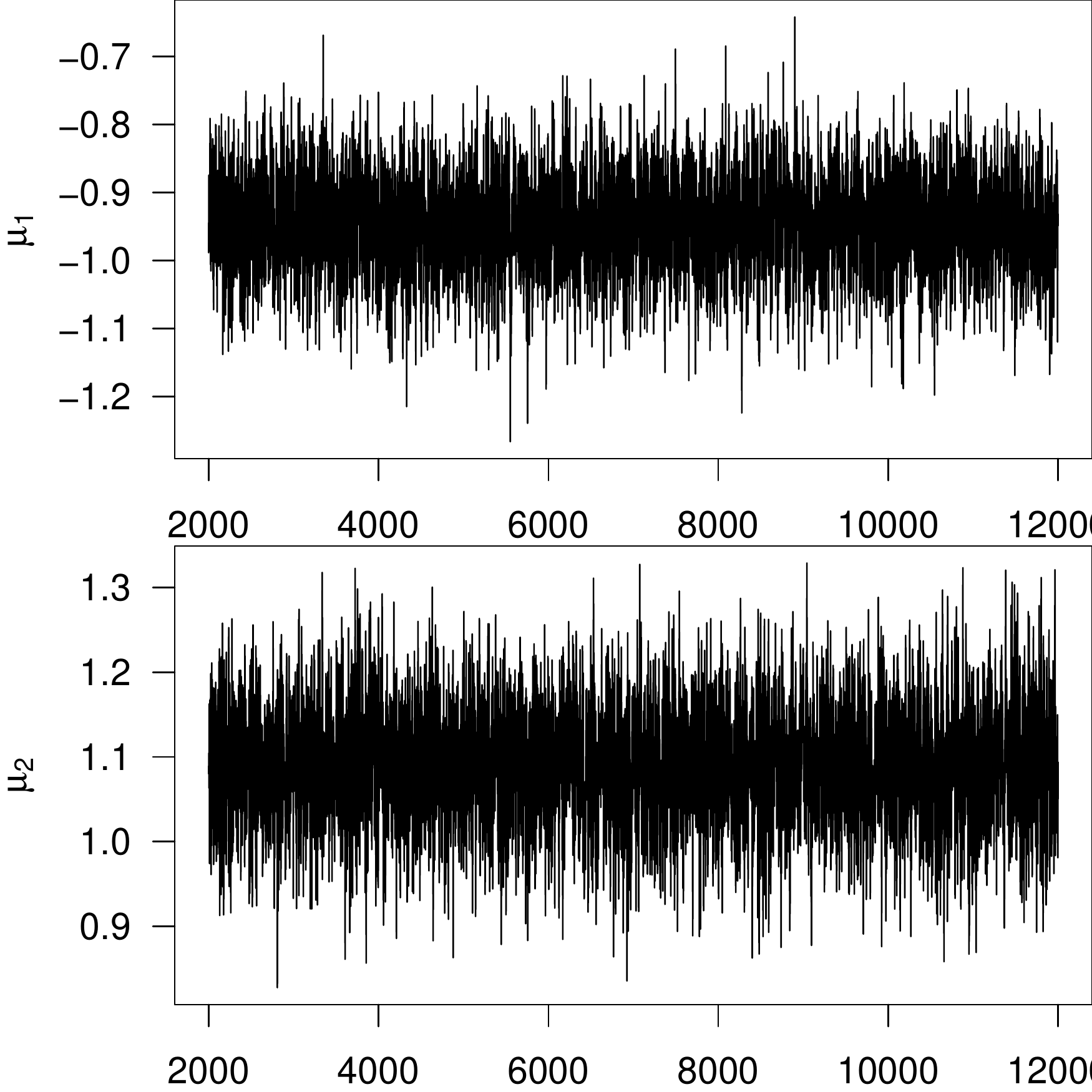}
&
\includegraphics[width=0.3\textwidth]{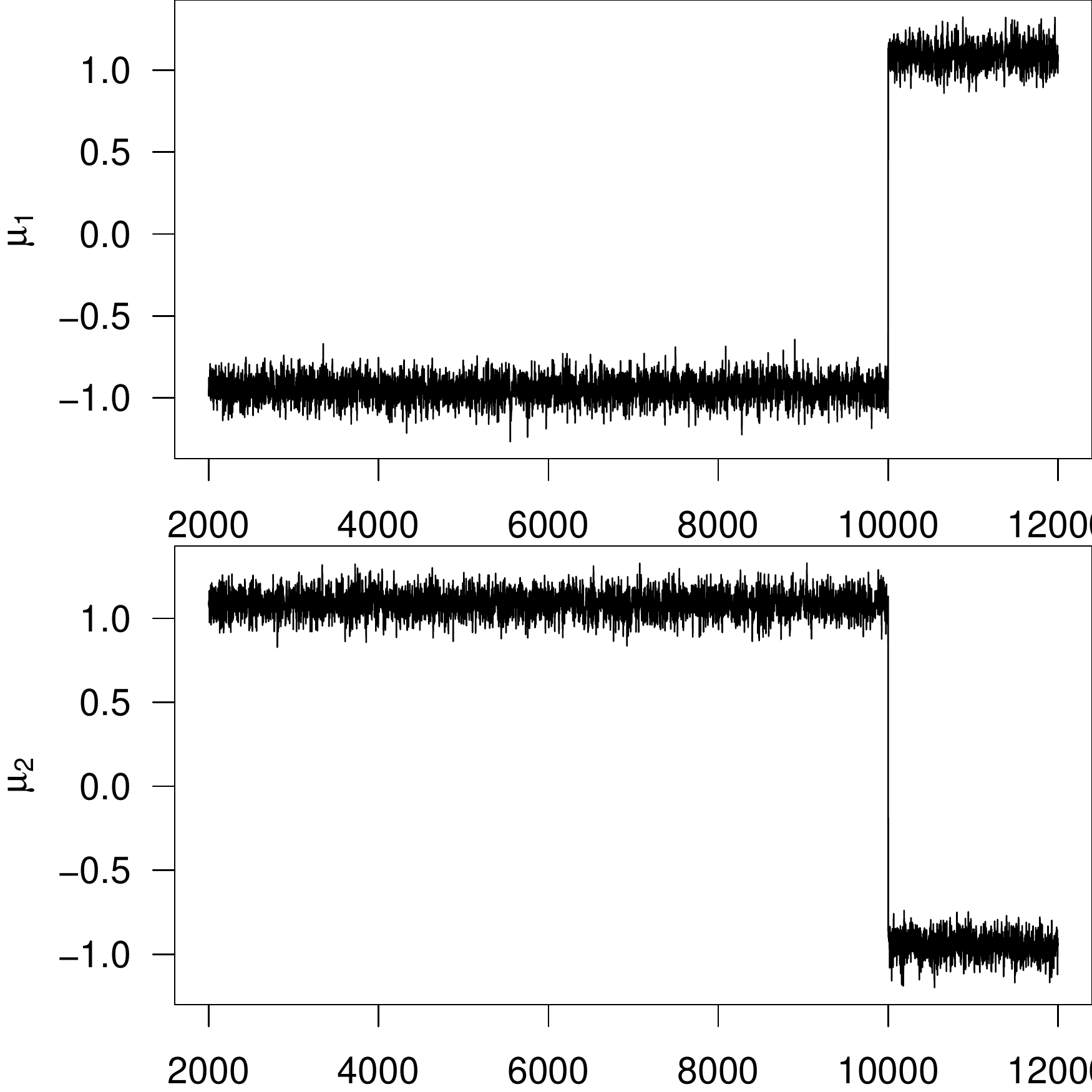}
\\
\includegraphics[width=0.3\textwidth]{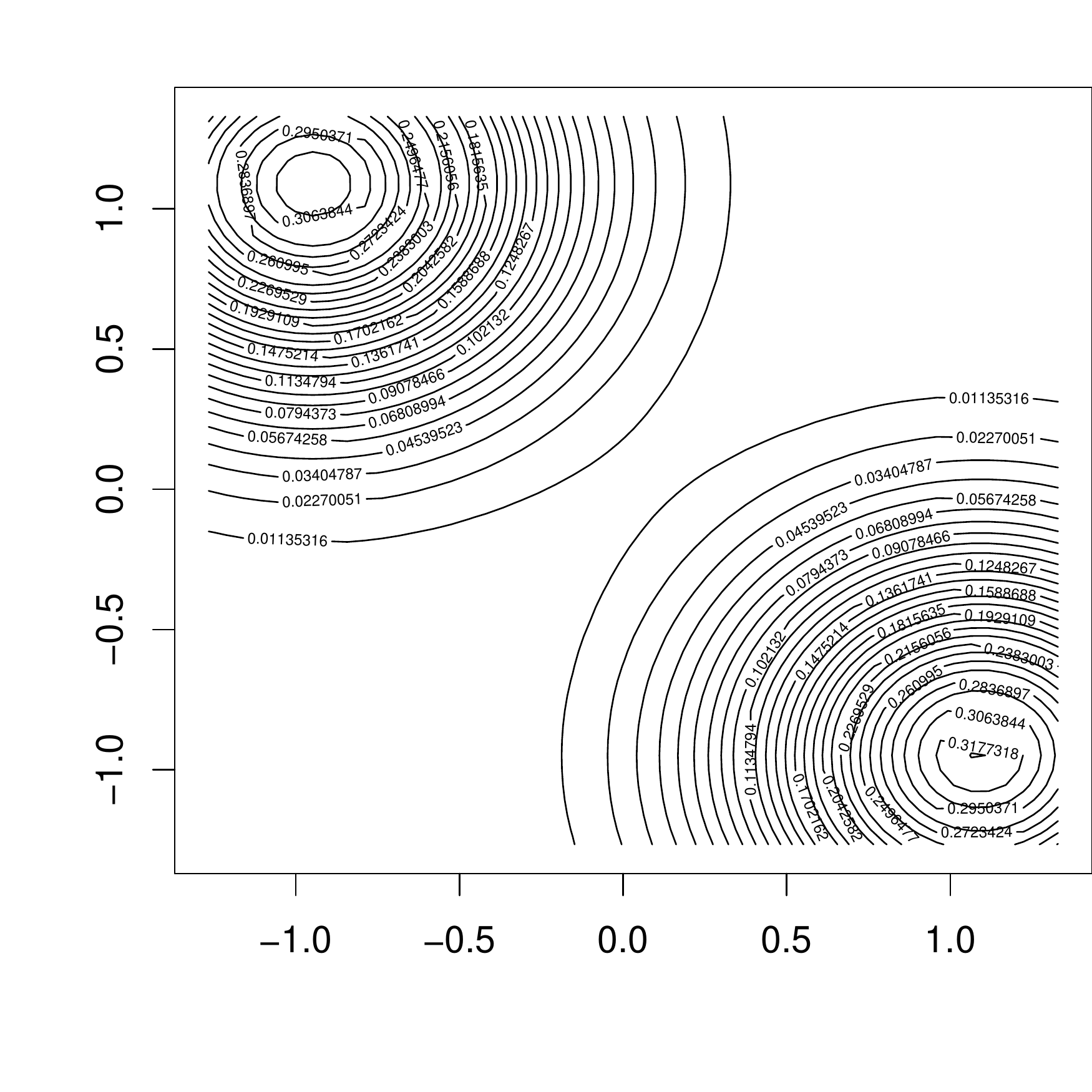}
&
\includegraphics[width=0.3\textwidth]{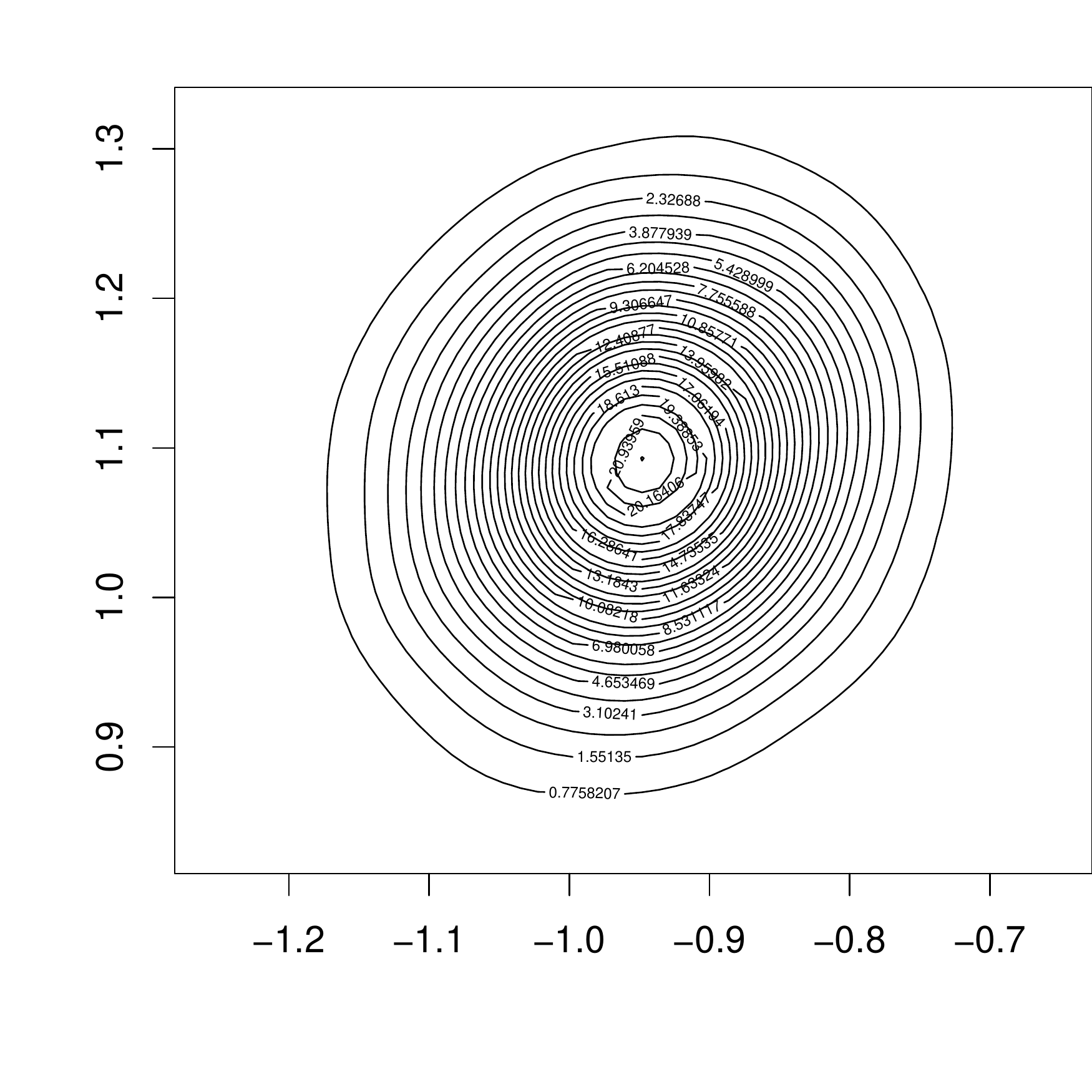}
&
\includegraphics[width=0.3\textwidth]{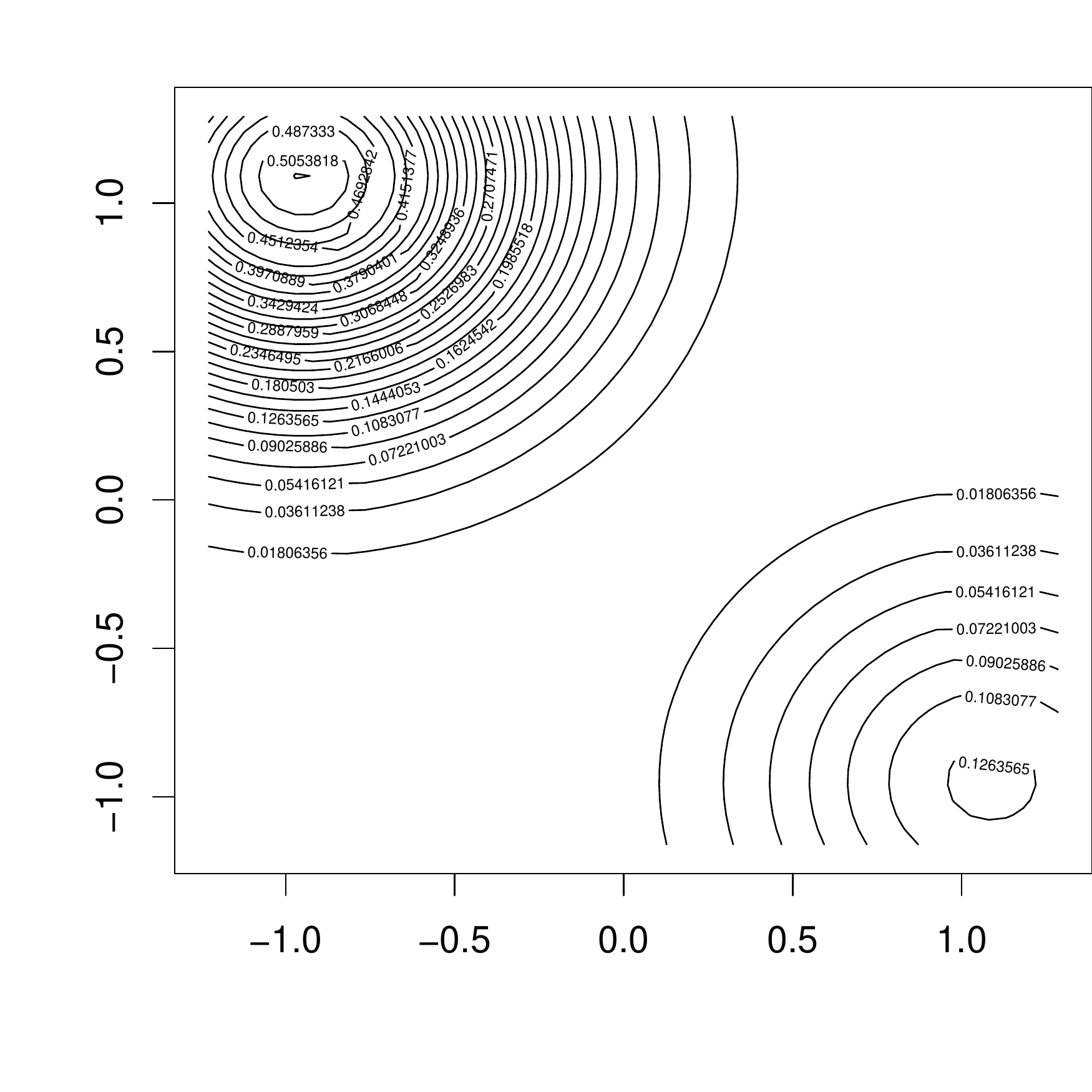}
\end{tabular}
\end{center}
\caption{\label{fig:mcmcexample}  MCMC chains for $\vmu$ (top row) and estimated posterior for $\vmu$ where 
{\it (a)} a perfect mixing occurs (each of the two permutations is visited with equal frequency); 
{\it (b)} no switching is exhibited;
{\it (c)} one random switch occurs.}
\end{figure}

In principle, a perfectly mixing chain should visit the points
$(\vmu,\vpi,\phi)$ and $(\vmu',\vpi',\phi)$ with the same frequency.
A chain with these characteristics for a model with $G=2$ and where $f(\cdot;\mu_g,\phi)$ is the Gaussian distribution with parameters $\mu_g$ and $\phi$, $\mathcal N(\mu_g,\phi)$, is depicted in Figure~\ref{fig:mcmcexample}{(a)}, together with the posterior distribution for $\vmu$.

A chain with a less than perfect mixing may either concentrate on one mode of the posterior distribution (Figure~\ref{fig:mcmcexample}{(b)}) or exhibit random switches (Figure~\ref{fig:mcmcexample}{(c)}). 

A naive, but effective, solution to the relabelling issue is to use a sampler which is inefficient with respect to the labelling -- that is, it is unlikely to switch labels -- but otherwise efficient \citep{puolamaki2009bayesian}. 
This can be an {\it ex post} solution, that is, we can ignore the relabelling issue if we verify that we obtained a chain where no switch occurred, but it is impractical in general terms since it is difficult to tune a sampler so that it is inefficient enough to avoid label switches but not too inefficient. 

\begin{figure}
\begin{center}
\begin{tabular}{ccc}
{\it (a)} & {\it (b)} & {\it (c)} \\
\includegraphics[width=0.3\textwidth]{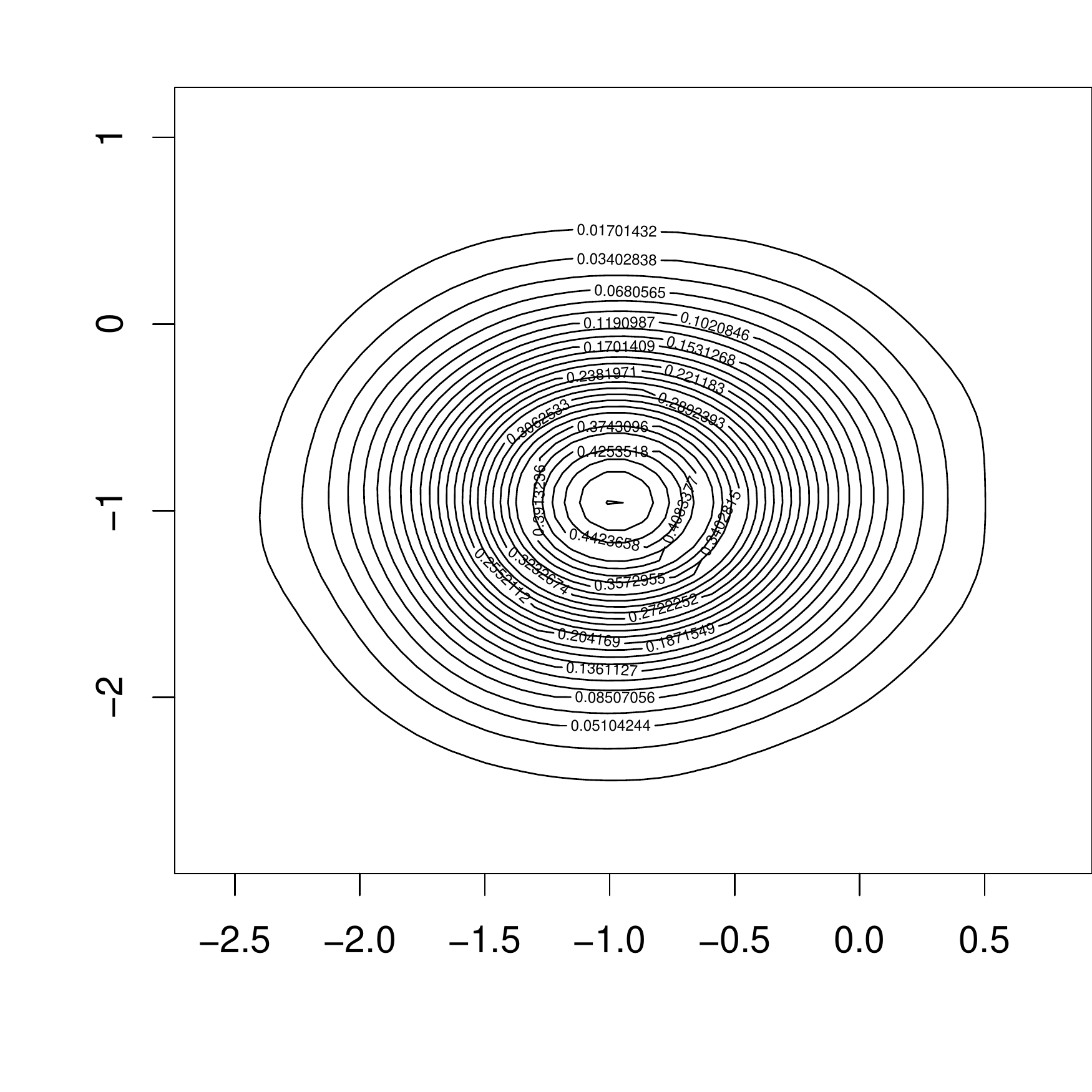}
&
\includegraphics[width=0.3\textwidth]{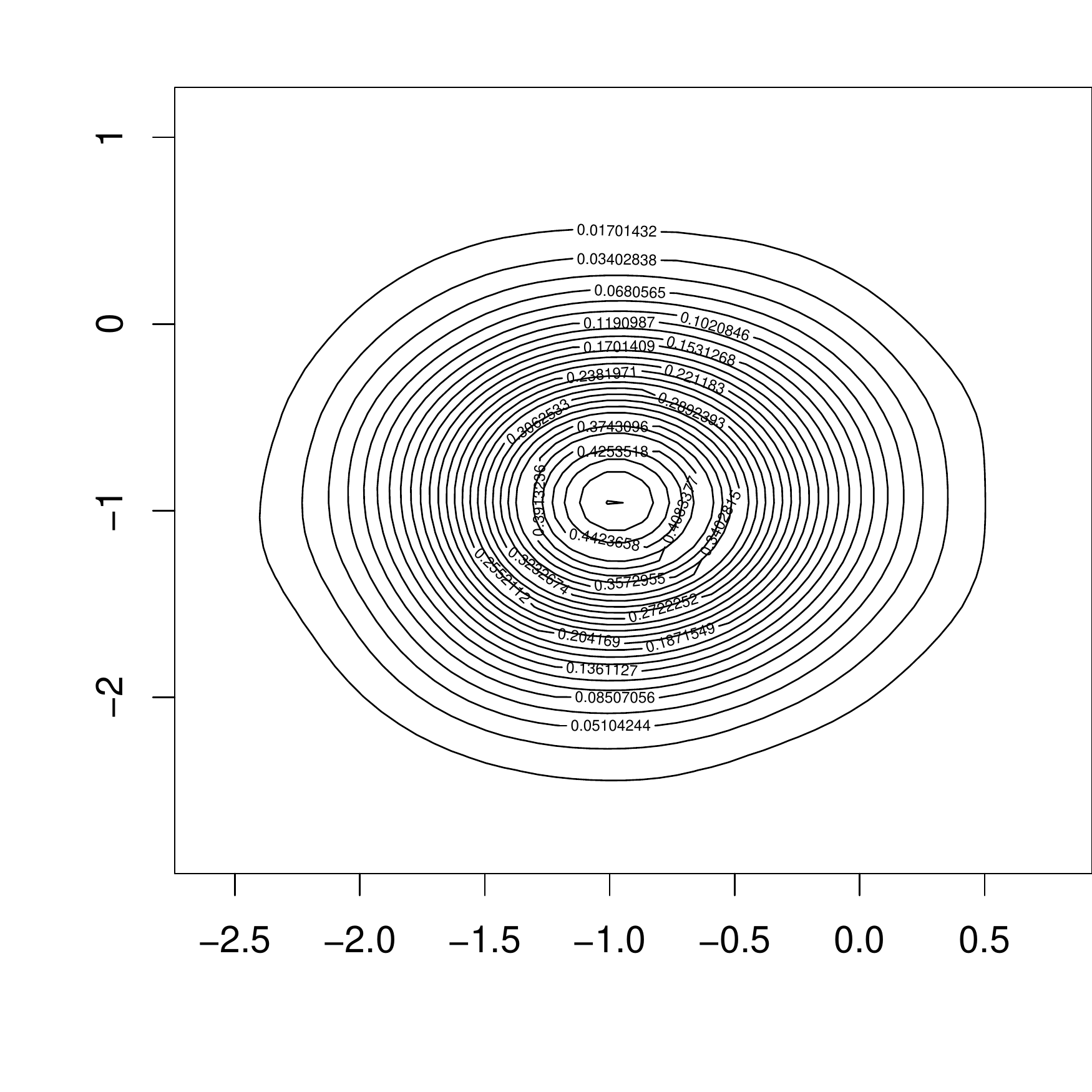}
&
\includegraphics[width=0.3\textwidth]{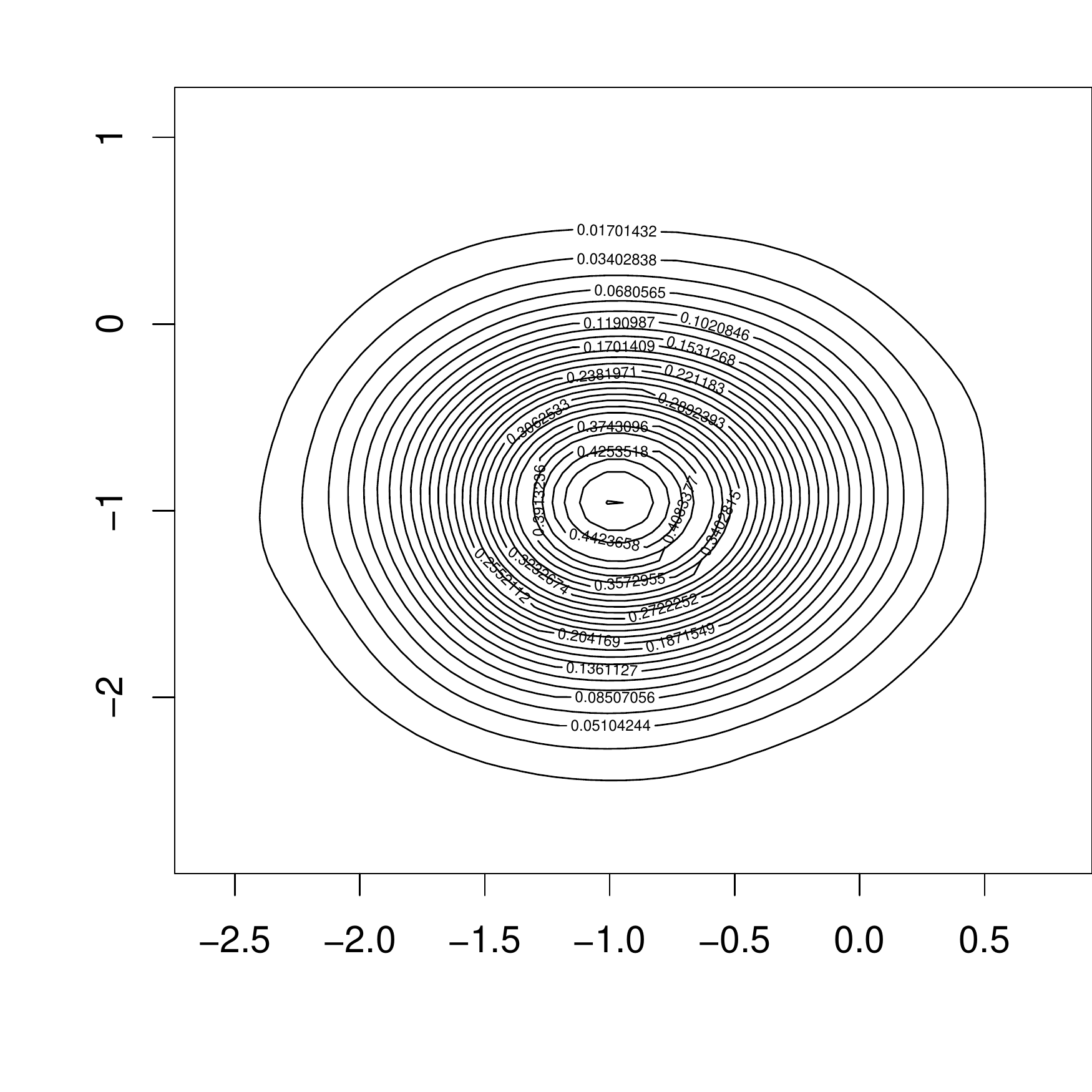}
\\
\includegraphics[width=0.3\textwidth]{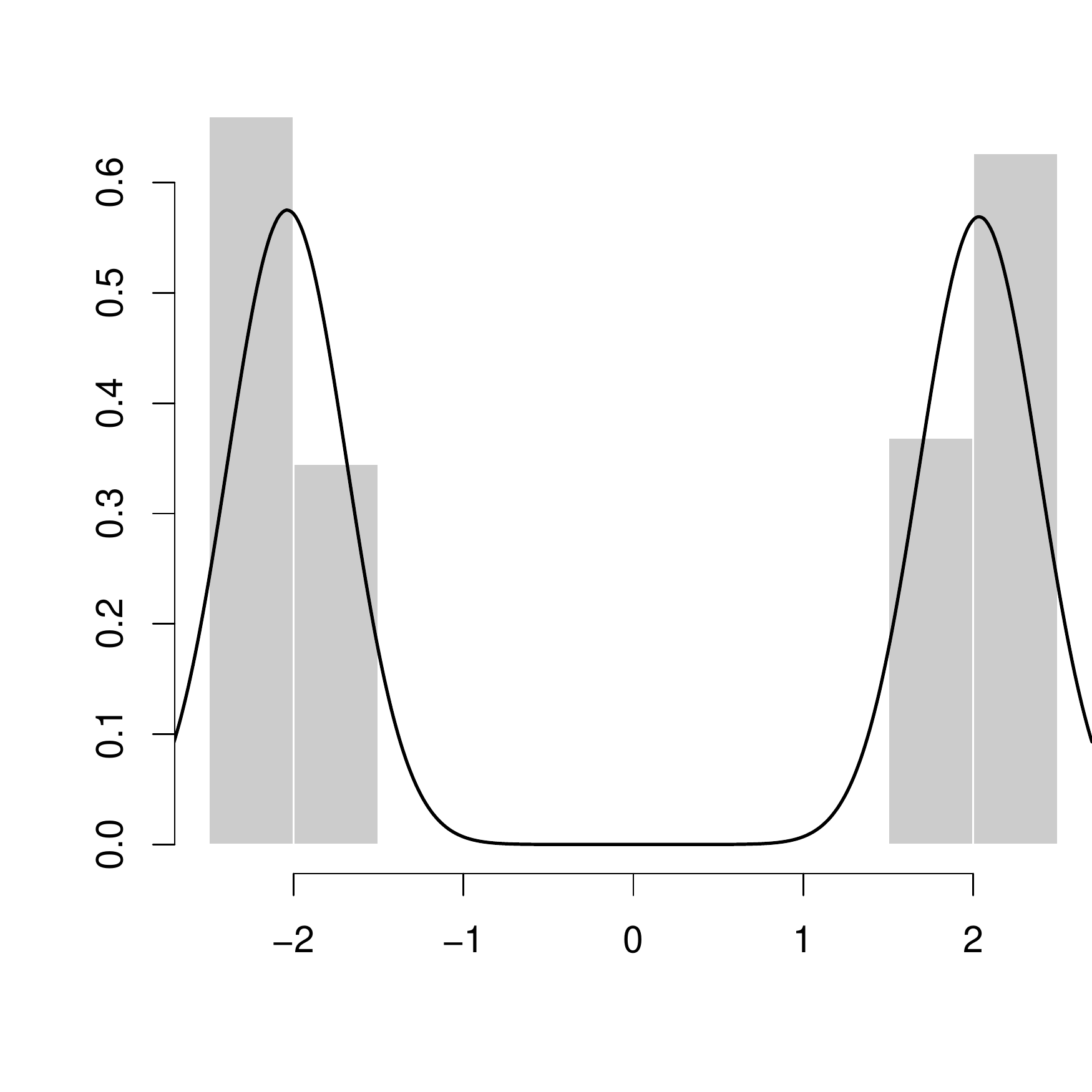}
&
\includegraphics[width=0.3\textwidth]{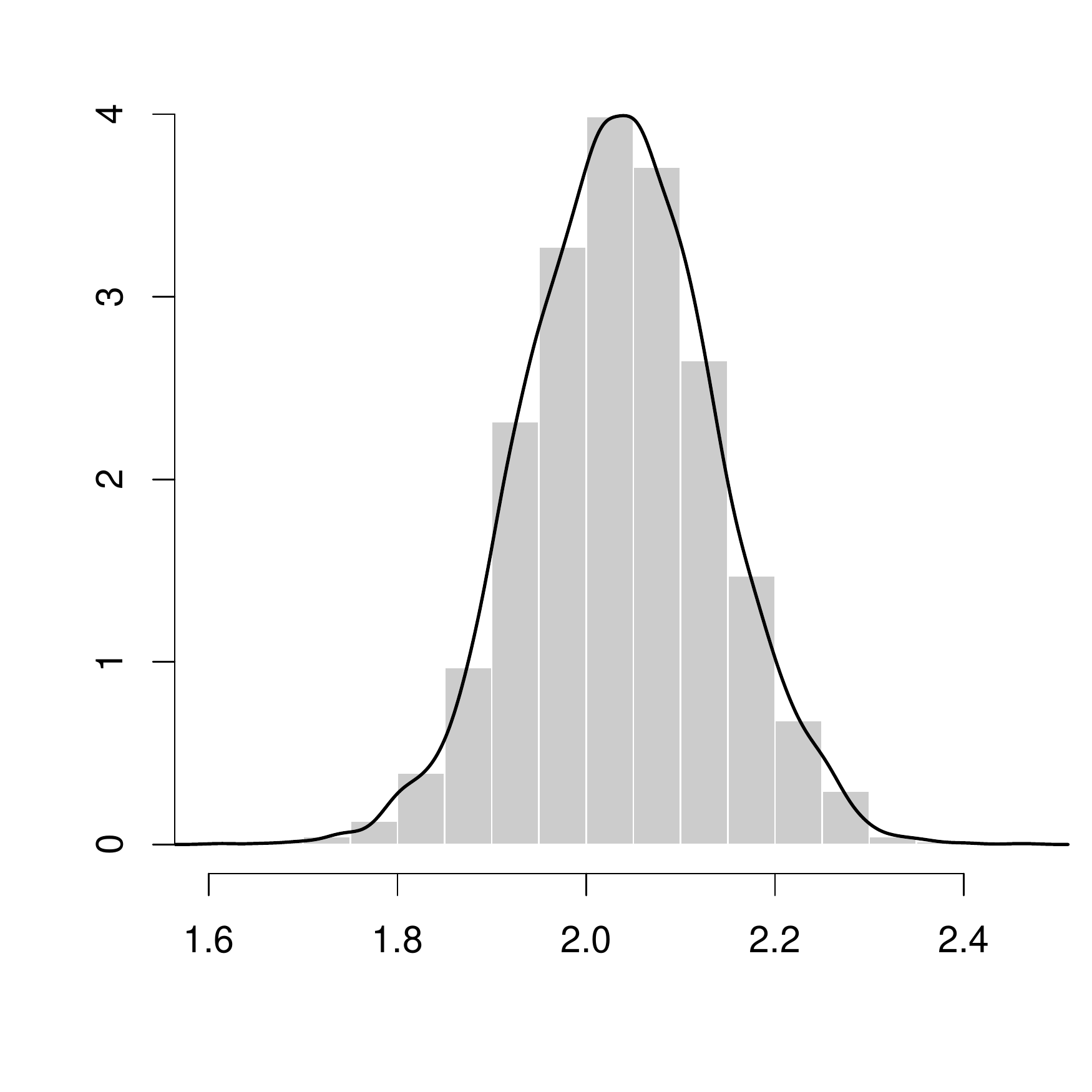}
&
\includegraphics[width=0.3\textwidth]{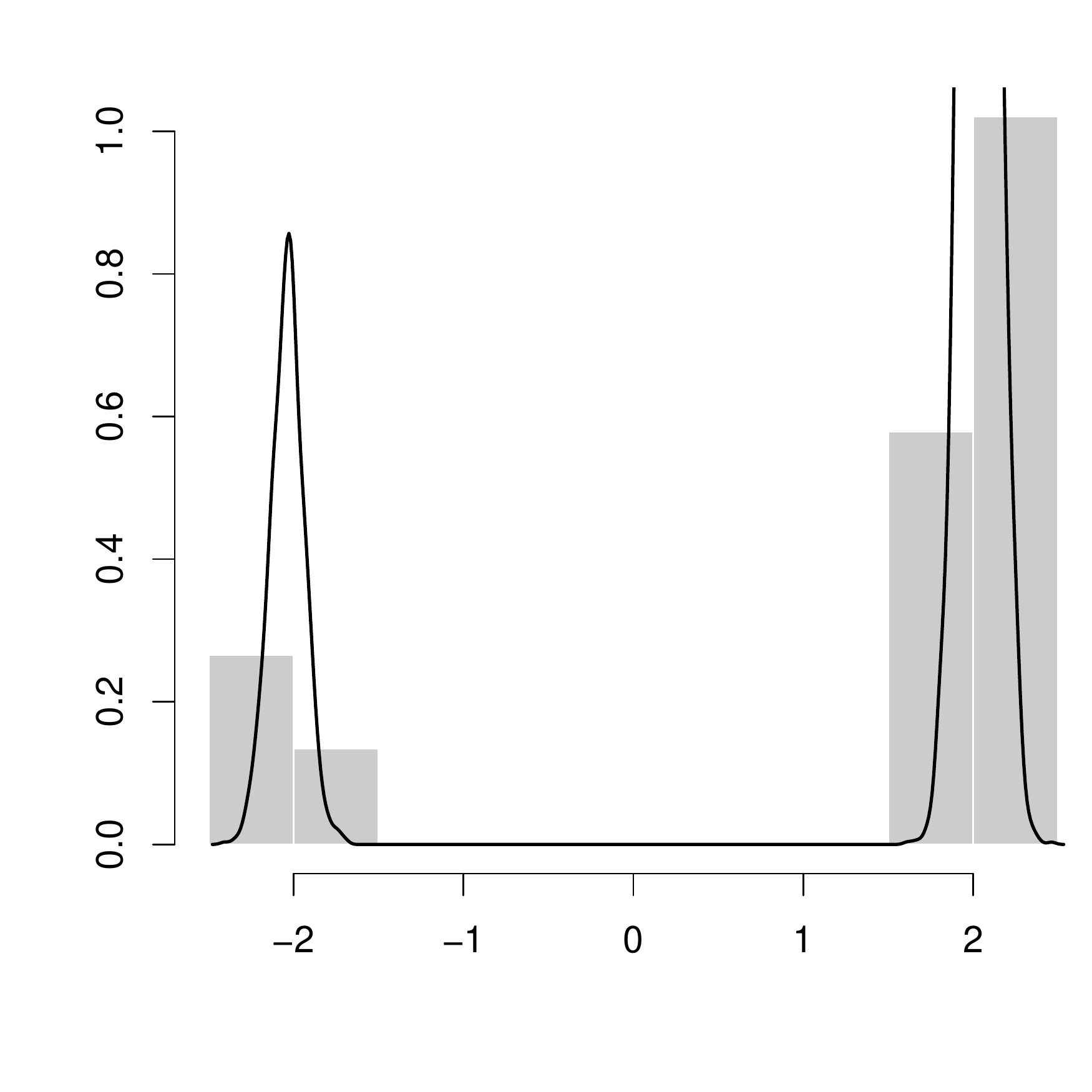}
\end{tabular}
\end{center}
\caption{\label{fig:mcmcexamplePost}  Estimated posteriors for $(y_1,y_2)$ (top row) and $\mu_2-\mu_1$ (bottom row) based on chain 
{\it (a)}, {\it (b)}, {\it (c)} of Figure~\ref{fig:mcmcexample}.}
\end{figure}

We note that the presence of label switches (or the whole issue of relabelling) is totally not relevant if the quantities we are interested in are invariant with respect to the labels, as is the case for a prediction of $(y_1,y_2)$ (depicted in Figure~\ref{fig:mcmcexamplePost}, top row), or the inference for the parameter $\phi$. 

A particularly relevant example of invariant quantity is the probability of two units being in the same group, $c_{ij}=P(Z_i=Z_j|\data)$, $i,j=1,\ldots,n$, whose estimate based on the sample is
\begin{equation}
 \hat{c}_{ij} = \frac{1}{H} \sum_{h=1}^H \uno{\ca{Z_i}=\ca{Z_j}}. 
 \label{eq:Cmatrix}
\end{equation}
The $n\times n$ matrix $C$ with elements $\hat{c}_{ij}$ can be seen as an estimated similarity matrix between units, and the complement to one $\hat{s}_{ij}=1-\hat{c}_{ij}$ as a dissimilarity matrix (note that it is not a distance metric as $s_{ij}=0$ does not imply that the units $i$ and $j$ are the same).

Relabelling becomes relevant when we are interested, directly or indirectly, in the features of the $G$ groups, for example the posterior (and predictive) distributions of component-related quantities such as the difference $\mu_2-\mu_1$ or the probability of each unit belonging to each group, $q_{ig}=P(Z_i=g|\data)$, whose MCMC estimate is
\begin{equation}
\hat{q}_{ig} = \frac{1}{H} \sum_{h=1}^H \uno{\ca{Z_i}=g},
\label{eq:probappartstimata}
\end{equation} 
for $i=1,\ldots,n$ and $g=1,\ldots,G$.

In Figure~\ref{fig:mcmcexamplePost}, bottom row, we depict the posterior distribution of $\mu_2-\mu_1$ based on the samples $\{\ca{\mu_2}-\ca{\mu_1}:h=1,\ldots,H\}$ obtained using the three chains. 
The first version is formally correct given that the model is not identified, but it is not able to tell us what is the average difference between the groups. 
The second version does answer to our question on the difference between the groups but is based on a very partial exploration of the posterior.
The third version leads to an incorrect answer.

It is then clear that the raw MCMC sample can not be used to study the posterior distributions of component-related quantities such as $\mu_g$ or $P(Z_i=g|\data)$. In order to study the posterior distributions of component-related quantities such as $\mu_g$, we need to define a suitable method to permute the labels at each iteration of the Markov chain. Then, the new labels are such that different labels do refer to different components of the mixture.

\section{Partitioning observations}
\label{sec:clustering}
A partition of the observations, meaning a point estimate of the group for each unit, can be easily obtained. 
Doing this, however, the issue of obtaining an estimate for groups features (posteriors of $\mu_g$) or the probability of units belonging to each group ($\hat{q}_{ig}$ in Equation~\eqref{eq:probappartstimata}) remains open.
In fact, the usual difficulties related to clustering techniques apply (for instance, the groups depend on the choice of the distance). A partition can be also obtained by maximizing the posterior distribution, notwithstanding the fact that the maximum is not unique (there are $G!$ modes), since the maxima are equivalent  any would be suitable.

Alternatively, the probabilities in Equation~\eqref{eq:Cmatrix} can be used to derive a partition of observations by employing some clustering technique based on a suitable similarity matrix.

A more sophisticated option, see \citet{fritsch2009improved}, involves defining a distance between partitions, for example
\begin{equation}
d(\vz^*,\vz)= \sum_{i<k} d_1\uno{z^*_i\neq z_k}\uno{z^*_i= z_k} + d_2\uno{z^*_i= z_k}\uno{z^*_i \neq z_k},
\label{eq:distclusterA}
\end{equation}
and then search for the partition which minimizes the expected distance with the true groups $\bar{\vz}$, which means, if $d_1=d_2=1$, find $\vz^*$ which minimizes
\begin{equation}
E(d(\vz^*,\bar{\vz})|\data) = \sum_{i<k}\big|\uno{z^*_i=z^*_k}-c_{ik}\big|,
\label{eq:mindist}
\end{equation}
where $c_{ik}$ can be replaced by $\hat{c}_{ik}$.

Alternative distances between partitions may be used, for instance the Rand index $d_2(\vz^*,\vz)=1-d(\vz^*,\vz)\left(\binom{n}{2}\right)^{-1}$ or the adjusted Rand index \citep{hubert1985comparing}.

Note that if the distance function is a linear operator then the following holds:
\begin{equation}
E(d(\vz^*,\vz)|\data) = d\left(\vz^*,E(\vz|\data)\right).
\label{eq:lineardistance}
\end{equation}
The expectations in \eqref{eq:mindist} or \eqref{eq:lineardistance} can be obtained using the MCMC sample as 
\begin{equation}
 E(d(\vz^*,\vz)|\data) = \frac{1}{H}\sum_{h=1}^H d(\vz^*,\ca{\vz}).
\label{eq:expdistmcmc}
\end{equation}
The optimization should be done in the space of all possible partitions, since this can be very large, the authors suggest performing optimization on a suitable subset, reasonable alternatives being the set $\{\ca{\vz}\}$ or the set of clusterings resulting from different classical algorithms applied to the similarity matrix \eqref{eq:Cmatrix} (or the union of the two).

\section{Obtaining probabilities of belonging to a group}
\label{sec:probgroups}
\citet{puolamaki2009bayesian} deal with the relabelling issue considering as an objective the $n\times G$ matrix with elements $q_{ig}=P(Z_i=g)$ ($\tilde{\beta}$ in their notation).
This is obtained by maximizing the Bernoulli likelihood.
The latter can be specified according to two alternative formulations.
The first is one in \citet{puolamaki2009bayesian}, where the $HG\times n$ matrix $Z'$ is such that
\[ Z'_{ri} = 1 \mbox{ iff } \ca{Z_i} = g \mbox{ where } r=G(h-1)+g, \]
and get
\begin{equation}
L = \prod_{r=1}^R\sum_{g=1}^G\prod_{i=1}^n q_{ig}^{Z'_{ri}}(1-q_{ig})^{1-Z'_{ri}}. 
\end{equation}
The above likelihood can also be written as
\begin{equation}
L= \sum_{h=1}^H \sum_{g=1}^G\prod_{i=1}^n q_{ig}^{\uno{\ca{Z_i}=g}}(1-q_{ig})^{1-\uno{\ca{Z_i}=g}}.
\end{equation}
The intuitive idea behind this strategy is that if two units $i_1$ and $i_2$ often belong to the same group, that is, $Z'_{r,i_1}=Z'_{r,i_2}$ for many $r$, then they should be assigned to the same group, thus leading to a high value of $q_{i_1\bar{g}}$ and $q_{i_2\bar{g}}$ for some value of $\bar{g}$. 
Note that the likelihood above is itself labelling invariant, thus it has $G!$ maxima. 

An EM algorithm is proposed to perform the optimization:
\begin{description}
\item[E step:] for each row $r$ (which represent a group in an iteration) and for each group obtain
\[
\gamma_{rg} = \frac{\prod_{i=1}^n q_{ig}^{Z'_{ri}}(1-q_{ig})^{1-Z'_{ri}}}%
{\sum_{g=1}^G \prod_{i=1}^n q_{ig}^{Z'_{ri}}(1-q_{ig})^{1-Z'_{ri}}} =
\frac{p((z_{1g},\ldots,z_{ng})|\theta)}{\sum_{g=1}Gp((z_{1g},\ldots,z_{ng})|\theta)};
\]
\item[M step:] compute the mean of the $Z'_{ri}$ with weights $\gamma_{rg}$
\[
q_{ig} = \frac{\sum_{r=1}^R\gamma_{rg}Z'_{ri}}{\sum_{r=1}^R\gamma_{rg}}
= \frac{\sum_{r:Z'_{ri}=1}\gamma_{rg}}{\sum_{r=1}^R\gamma_{rg}}.
\]
\end{description}

Equivalently, the matrix $Q$ can be found minimizing the cost function
\[ \prod_{h=1}^H \sum_{\nu\in\mathcal V}\frac{1}{G!} q_{i\nu(\ca{Z_i})}. \]

\section{Relabeling methods}
\label{sec:relabeling}
Relabelling means permuting the labels at each iteration of the Markov chain in such a way that the relabelled chain can be used to draw inference on component specific parameters. Loosely speaking we may say that the relabelled chain can be seen as a chain where no label switching has occurred or, in other words, the new labels are such that different labels do refer to different components of the mixture.

One method to perform the relabelling involves imposing identifiability constraints such as $\pi_1<\pi_2<\ldots<\pi_G$ or $\mu_1<\mu_2<\ldots<\mu_G$.
Equivalently, this may be seen as a conditioning of the full (multimodal) posterior where the conditioning event is the identifiability constraint.
Such a solution, although theoretically sound, may not be applicable when an obvious constraint does not exist and it may not work well if the components are not well separated \citep{stephens2000dealing,jasra2005markov}. 


It is worth noting that relabelling strategies may act during the MCMC sampling, and/or they may be used to post-process the chains. 
In general, those solutions which post-process the chains are particularly convenient (since the issue can be ignored in performing the MCMC and then dealt with later). 
Generally, existing relabelling procedures select the permutation of the labels that minimizes a well defined distance between some components, such as pivots and classification probabilities, at each MCMC iteration. \citet{papastamoulis2016label} provides the \textbf{label.switching} \texttt{R} package with a range of deterministic and probabilistic methods for performing relabelling: in Section~\ref{sim:study} and \ref{case:study} a comparison between some of these alternatives and our methodology will be provided. 

\subsection{Decision theoretic approach}
A rather general decision theoretic framework for the relabelling problem is proposed by \citet{stephens2000dealing}. Such approach translates the problem to that of choosing an action $a$ from a set of actions $\mathcal A$ where a loss function $\mathcal L:\mathcal A\times\Theta\rightarrow\mathbb R$ represents the loss we incur if we choose the action $a$ and the true value of the parameter is $\theta$.

The loss function makes sense if it is permutation invariant (remember that if we permute the parameter the model remains the same), we can obtain a permutation invariant loss function $\mathcal L$ from a non invariant one $\mathcal L_0$ by defining
\[ \mathcal L(a;\theta) = \underset{\nu}{\min} \mathcal L_0(a;\nu(\theta)). \]

The action $a$ is then chosen by minimizing the posterior expected loss
\[ \mathcal R(a) = E(\mathcal L(a;\theta)|\data), \]
which can be approximated using the MCMC sample by
\begin{eqnarray}
\hat{\mathcal R}(a) 
&=& \frac{1}{H}\sum_{h=1}^H \mathcal L(a;\ca{\theta})  \\
&=& \frac{1}{H}\sum_{h=1}^H \underset{\nu_h}{\min} \mathcal L_0(a;\nu_h(\ca{\theta}))  \\
&=&\underset{\nu_1,\ldots,\nu_H}{\min} \left( \frac{1}{H}\sum_{h=1}^H  \mathcal L_0(a;\nu_h(\ca{\theta})) \right).
\end{eqnarray}
The action $a$ can be the estimation of the parameter (or part of it) and the loss function may be a distribution to be fitted or an estimation error, the choice should be driven by the objective of inference.
If the objective is the clustering of $n$ units into $G$ groups a reasonable action is reporting the $n\times G$ matrix
$Q = [q_{ig}]$
where $q_{ig}$ is the probability that the $i$-th unit belongs to the group $g$.
A corresponding loss is then the distance, somehow measured (\citet{stephens2000dealing} employs the Kullback-Leibler distance), between $Q$ and its true value $P(\theta)=[p_{ig}(\theta)]$ where (for the toy example)
\[ p_{ig}(\theta)= P(Z_i=g|\vy,\theta) = \frac{\pi_g f(y_i;\mu_g,\theta)}{\sum_j \pi_j f(y_i;\mu_j,\theta)} \]

The general algorithm for performing \citet{stephens2000dealing} method is as follows
\begin{description}
\item[Start:] from arbitrary permutations $\nu_1,\ldots,\nu_H$.
\item[Step 1:] obtain $a=\underset{a}{\mbox{argmin}} \sum_{h=1}^H  \mathcal L_0(a;\nu_h(\ca{\theta}))$.
\item[Step 2:] obtain $\nu_h=\underset{\nu_h}{\mbox{argmin}} L_0(a;\nu_h(\ca{\theta}))$.
\end{description}
Note that step 2 entails $n$ minimizations with respect to all the permutations ($G!$), \citet{stephens2000dealing} points out the existence of efficient numerical algorithm if the loss function $\mathcal L_0$ can be written as $\mathcal L_0=\sum_{g=1}^G \mathcal L_0^{(g)}(a;\pi_g,\mu_g,\phi)$.\

A problem with this method might be the choice of the appropriate loss function/the dependence of the results on the loss function. Our method presented in Sect.~\ref{sec:pivotal} does not require a minimization step, and for this reason might be computationally appealing in many situations.

\section{Pivotal method}
\label{sec:pivotal}
Suppose that a partition of the observations in $\hat{G}$ groups, $\mathcal G_1,\ldots,\mathcal G_{\hat{G}}$ has been obtained as discussed in Section~\ref{sec:clustering}.
As already pointed out, this may be enough for some purposes, but we may be interested in the probabilities $P(Z_i=g)$ and in the posteriors for groups parameters, $\mu_g$.

Suppose that we can find $\hat{G}$ units, $i_1,\ldots,i_{\hat{G}}$, one for each group, which are (pairwise) separated with (posterior) probability one (that is, the posterior probability of any two of them being in the same group is zero). In terms of the matrix $C$, the $\hat{G}\times \hat{G}$ sub-matrix with only the row and columns corresponding to $i_1,\ldots,i_{\hat{G}}$ will be the identity matrix.

We then use the $\hat{G}$ units, called pivots in what follows, to identify the groups and to relabel the chains: for each $h=1,\ldots, H$ and $g=1,\ldots,\hat{G}$
\begin{equation}
 \ca{\mu_g}=\ca{\mu_{\ca{Z_{i_g}}}};
\label{eq:relabelmu}
\end{equation}
\begin{equation}
 \ca{Z_i}=g \mbox{ for } i:\ca{Z_i}=\ca{Z_{i_g}}.
\label{eq:relabelZ}
\end{equation}

The availability of $\hat{G}$ perfectly separated units is crucial to the procedure, and it can not always be guaranteed.
We now discuss three different circumstances under which the relabelling procedure is unsuitable
\begin{description}
\item[(i)] the number of actual groups in the MCMC sample is higher than $\hat{G}$;
\item[(ii)] the number of actual groups in the MCMC sample  is lower than $\hat{G}$;
\item[(iii)] the number of actual groups in the MCMC sample  is equal to $\hat{G}$ but the pivots are not perfectly separated.
\end{description}

Let us first clarify what is meant by the number of actual groups. 
The model has $G$ components, but some mixture components may be empty in the Markov chain, that is, it may happen that  $\#\{g: \ca{Z_i}=g\mbox{ for some }i\}<G\, \forall h$. By actual number of groups we mean the number of non empty groups, $G_0$ in what follows.
It is then clear that the Markov chain does not have informations on more than $G_0$ groups.

We also note that the number of non empty groups may vary with iterations, let
\[
\ca{G} = \#\{g: \ca{Z_i}=g\mbox{ for some }i\}.
\]

\newcommand{\hh}{\mathcal H}

Consider now the set $\hh_1\subset\{1,\ldots,H\}$ of iterations where $\ca{G}>\hat{G}$; some units and groups will then have no available pivot.
These units will not be attributed any group by performing (\ref{eq:relabelZ}). Thus for these units
\[ \sum_{g=1}^{\hat{G}} \hat{P}(Z_i=g)
=\sum_{g=1}^{\hat{G}}\hat{q}_{ig} 
= \sum_{g=1}^{\hat{G}}\frac{1}{H} \sum_{i=1}^H \uno{\ca{Z_i}=g}
 < 1. \]
We suggest cancelling those iterations of the chains where this occur, that is, the final --partial-- chain is a sample from the posterior conditional on having at most $\hat{G}$ non empty groups.

Consider now the set $\hh_2\subset\{1,\ldots,H\}$ of iterations where $\ca{G}<\hat{G}$; if  $h\in\hh_2$, $\ca{Z_{i_k}}=\ca{Z_{i_s}}$ for some pivots $i_k, i_s$.
As a consequence, $\hat{c}_{hk}>0$: the pivots are not perfectly separated. 
The procedure in \eqref{eq:relabelmu} and \eqref{eq:relabelZ} can not be performed (it is not well defined), so also in this case we will have to cancel the corresponding part of the chain. Finally, consider the set 
\[ 
\hh_3=\{ h : \exists k,s \mbox{ s.t. } \ca{Z_{i_k}}=\ca{Z_{i_s}} \}
\]
that is, the set of iterations where (at least) two pivots are put in the same group. Note that $\hh_2\subset \hh_3$ but $\hh_3$ may be larger.
The same provision as above applies, we need to get rid of this part of the chain.
In the end, we will relabel the chain with iterations 
\begin{equation}
 \hh_0 = \{1,\ldots, H\} \backslash (\hh_1\cup\hh_2\cup\hh_3)
\label{eq:itvalid}
\end{equation}
which can be considered a sample from the posterior distribution conditional on (i) there being exactly $\hat{G}$ non empty groups, (ii) the pivots falling into different groups. 
%
%
\subsection{Pivots identification}
\label{piv:id}

A relevant issue is how to identify the pivots, noting that perfectly separated  units may not exist and that, even if they exist, we may not be able to find them since the set of all possible choices is too big to be fully searched.

The general method we put forward is to select a unit for each group according to some  criterion, for instance for group $g$ containing units $\mathcal G_g$ we may chose $\bar{i}\in\mathcal G_g$ that maximizes one of the quantities
\begin{equation}
\label{eq:maxmeth}
  \underset{j\in\mathcal G_g}{\max}\, c_{\bar{i}j},  \;\;\;\;
  \sum_{j\in\mathcal G_g} c_{\bar{i}j},  \;\;\;\;
  \sum_{j\in\mathcal G_g} c_{\bar{i}j} - \sum_{j\not\in\mathcal G_g} c_{\bar{i}j}; 
\end{equation}
or minimizes one of the quantities
\begin{equation}
\label{eq:minmeth}
 \underset{j\in\mathcal G_g}{\min}\, c_{\bar{i}j},  \;\;\;\;
  \underset{j\not\in\mathcal G_g}\, c_{\bar{i}j},  \;\;\;\;
  \sum_{j\not\in\mathcal G_g} c_{\bar{i}j}.  
\end{equation}

We introduce a further method, which we call \textit{Maxima Units Search} (hereafter MUS), that turns out to be suitable in case of a low number of mixture component, e.g., $G=3, 4$. This procedure differs from the others in the strategy for detecting pivots, since it does not rely upon a
maximization/minimization step but it identifies those units satisfying a proper search within the estimated similarity matrix $C$
(see the Appendix for more details on the MUS procedure).

The quality of the choice of pivotal units by the proposed methods is measured by the probability of the conditioning event
\begin{equation}
 \frac{1}{H} \#\hh_0.
 \label{eq:condevent}
 \end{equation}
estimated by the (original, raw) MCMC sample. It is worth noting that the idea of solving the relabelling issue by fixing the group for some units dates back to \citet{chung2004difficulties}, who, however, gave no indication on how to choose the units. Also, since they suggest imposing such a restriction in the MCMC, there is no measure of the extent to which it influences the result (of the extent to which it is informative if we interpret it as a prior information).
We note, however, that \citet{chung2004difficulties} may be very interesting when a set of units which are to be attributed to different groups can be defined exogenously.

Another related idea is put forward by \citet{yao2014online}, who propose finding a reference labelling, that is, a clustering for the sample (for example, the posterior mode), and then relabel each iteration by minimizing some distance from the reference labelling.
The general idea is similar to the one we suggest, but it is more computationally demanding because of the required minimizations, on the other hand it avoids the need to condition on the pivots being separated. We can argue, however, that the latter is not a big drawback of our proposal since its effects can be measured and is likely to be small in many practical instances.

\section{Simulation study}
\label{sim:study}
The aim of this section is to evaluate the performance of the pivotal method introduced before. In particular, our goal is to investigate the behaviour of the proposed solution for dealing with label switching in different simulated scenarios. For this purpose, we focus on data simulated from a mixture of non-equally weighted mixtures of bivariate Gaussian distributions with unequal covariance matrices, so that the generated components may result in overlapping clusters.
Specifically, the simulation scheme consists in the following steps.

%
\begin{enumerate}
	\item[(i)] Simulate $n$ values $Y_{1}, \dots, Y_{n}$, from a mixture of mixtures of bivariate Gaussian distributions, where 
\begin{equation}
 \label{mixt:of:mixt}
 (Y_i|Z_i=g) \thicksim \sum_{s=1}^{2} p_{gs}\, \mathcal{N}_{2}(\mathbf{\mu}_{gs}, \Sigma_{s}).
 \end{equation}
That is, conditionally on being in group $g \in \{1,\dots,G\}$, $\mathbf{y}_{i}$ is picked out from one of two possible Gaussian  distributions 
with weights, means and covariances $p_{gs}$, $\mu_{gs}$ and $\Sigma_{s}$, $s=1,2$, respectively.  The likelihood of the model is then

\begin{equation*}
L(\vy;\vmu,\vpi,\Sigma) = \prod_{i=1}^n \sum_{g=1}^G \pi_g \left(\sum_{s=1}^{2} p_{gs} \, \mathcal{N}_{2}(\mathbf{\mu}_{gs}, \Sigma_s)\right).
\label{eq:lik:mixtofmixt}
\end{equation*}
	\item[(ii)] Obtain an MCMC sample which effectively explores all modes of the posterior distribution.
\item[(iii)] Estimate the $n\times n$ similarity matrix $C$ with elements $c_{ij}=P(Z_i=Z_j|\data)$, $i,j=1,\ldots,n$, by Equation~\eqref{eq:Cmatrix}.
\item[(iv)]	Apply a suitable clustering technique based on the estimated dissimilarity matrix with elements $\hat{s}_{ij}=1-\hat{c}_{ij}$ and obtain a partition of the observations in $\hat{G}$ groups with units $\mathcal{G}_{g}, g=1,\dots,\hat{G}.$
\item[(v)] Detect the pivots, one for each group, according to one criterion among the ones discussed before. 

	\item[(vi)] If necessary, discard those iterations of the chains belonging to $\hh_1\cup\hh_2\cup\hh_3$ (see Section~\ref{sec:pivotal}) and relabel the resulting chain with iterations in $\mathcal{H}_{0}$ (see Equation~\eqref{eq:itvalid}) via \eqref{eq:relabelmu} and \eqref{eq:relabelZ}.
\end{enumerate}

\begin{figure}
 \centering
 \subfloat[Scenario A]
 {\includegraphics[scale=0.31]{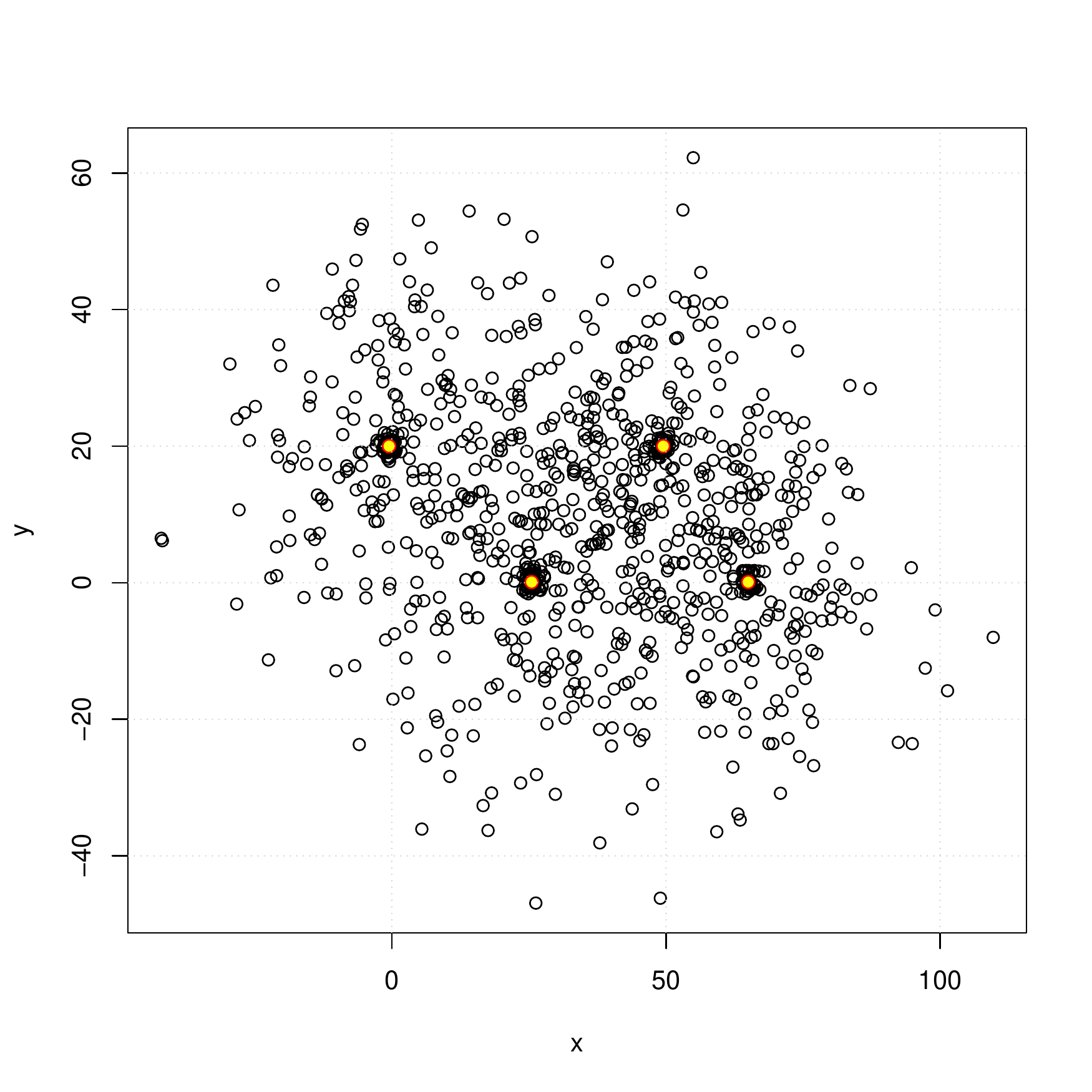}}~
 \subfloat[Scenario B]
{\includegraphics[scale=0.31]{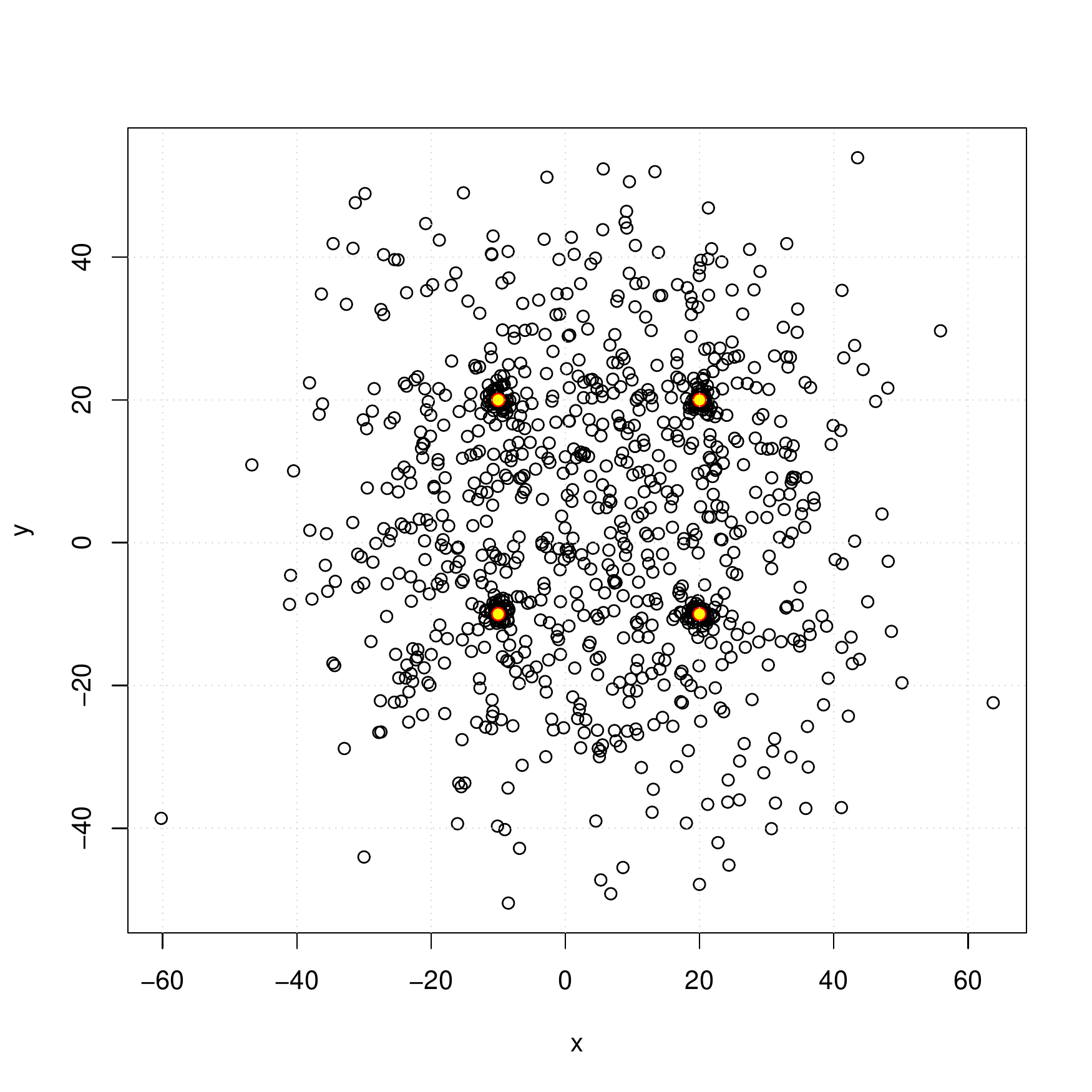}}~
\subfloat[Scenario C]
{\includegraphics[scale=0.31]{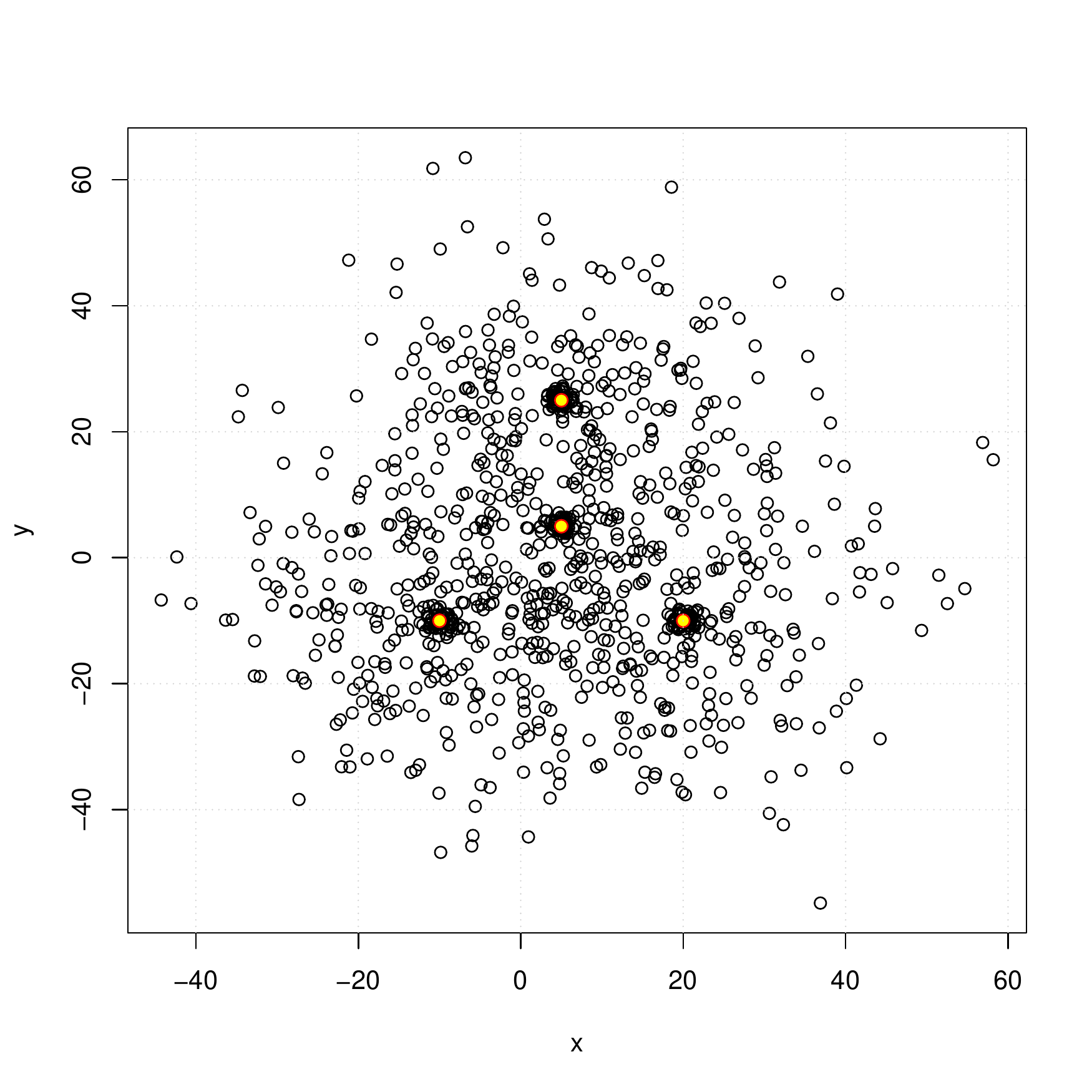}}
\vspace{-.2em}
\caption{Illustration of a simulated sample of size $n=1000$ from model (\ref{mixt:of:mixt}) with $G=4$ components, according to three different scenarios. The input means coordinates are reported in Table~\ref{tab:scenarios}.}
	\label{sim:data}
\end{figure}

In the following, a sample size of $n=1000$ and $G=4$ components are considered. For $g=1,\dots,4$, we set $\pi_{g}=1/4$,  $p_{g1}=0.2$, $p_{g2}=0.8$, and $\Sigma_{1}=\mathbf{I}_{2}$, $\Sigma_{2}=200\,\mathbf{I}_{2}$, being $\mathbf{I}_{2}$ the $2\times2$ identity matrix. We generate our simulated data from model~\eqref{mixt:of:mixt} (see Figure~\ref{sim:data}) using the input means reported in Table~\ref{tab:scenarios} and obtain an MCMC sample by considering $H=3000$ iterations. 

We proceed following points (i)-(vi) described above. As a remark, two different clustering strategies are applied on the dissimilarities $\hat{s}_{ij}$ in order to obtain $\hat{G}$ clusters of observations, namely the agglomerative and partitioning hierarchical clustering. Both methods only require a distance or a dissimilarity matrix as input and return a set of nested clusters that are organized as a tree. The former starts with the points as individual clusters and, at each step, merges the closest pair of clusters, according to some criterion to compute cluster proximity; the latter starts with one, all-inclusive cluster and, at each step, splits a cluster until only singleton clusters of individual points remain.\\
 We observe that the two algorithms provide very similar results in terms of the resulting clusters, and do not affect the performance of the relabelling procedure. Therefore, for the sake of illustration, we restrict to agglomerative hierarchical clustering, where the so-called \emph{complete linkage} is adopted as a common criterion for the computation of the dissimilarity between two clusters, since it is less susceptible to noise and outliers than other linkages.

\begin{table}
\centering
\begin{tabular}{rccc}
\hline
 & Scenario A & Scenario B & Scenario C \\
\hline
$\mu_{1s}$ & (25,0) & (-10,-10) &  (-10,-10) \\
$\mu_{2s}$ &  (60,0) & (20,-10) & (20,-10) \\
$\mu_{3s}$ & (0,20) & (-10,20) & (5,5) \\
$\mu_{4s}$ & (50,20) & (20,20) & (5,25)  \\
\hline
\end{tabular}
\caption{Two-dimensional mean vectors used as input for the three illustrated scenarios A, B and C, with number of mixture components $G=4$.}
\label{tab:scenarios}
\end{table}

\begin{figure}[H]
 \centering
 \subfloat[$ \underset{\bar{i}}{\max}\left(\max_{j \in \mathcal{G}_{g}} c_{\bar{i}j}\right)$] 
{\includegraphics[width=.33\textwidth]{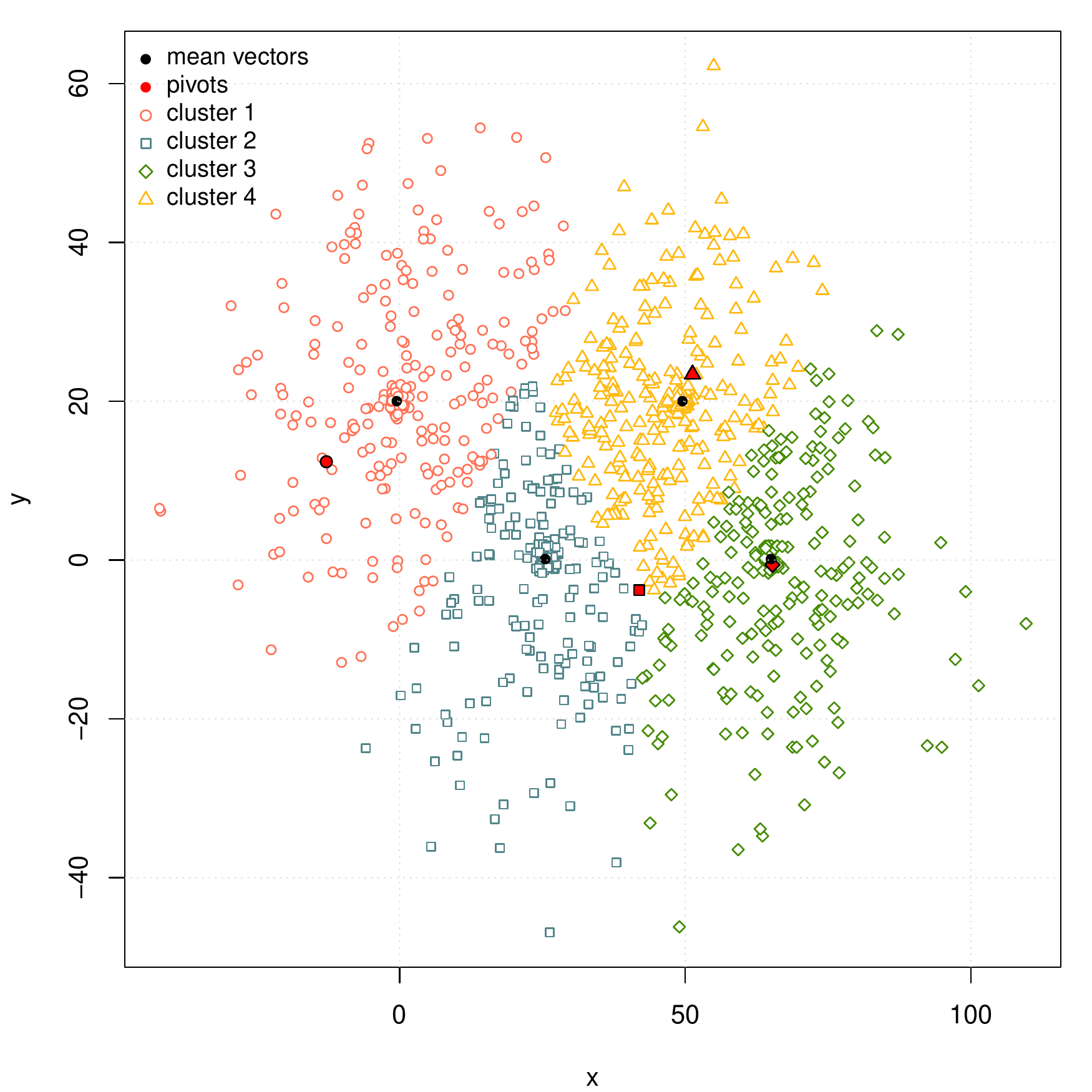}}~
\subfloat[$\underset{\bar{i}}{\max}\sum_{j \in \mathcal{G}_{g}} c_{\bar{i}j} $]
{\includegraphics[width=.33\textwidth]{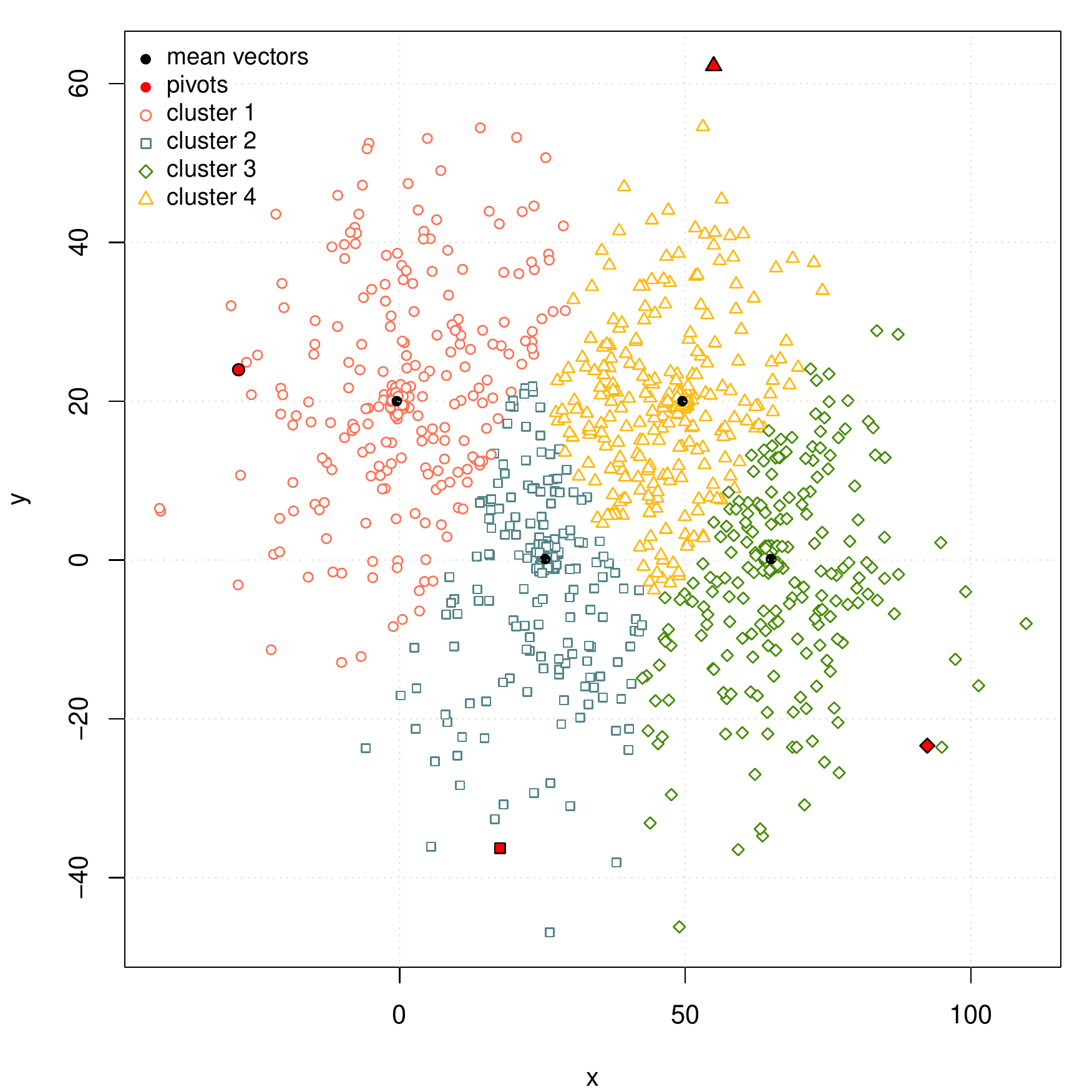}}~
 \subfloat[$\underset{\bar{i}}{\min}(\underset{j\in\mathcal G_g}{\min}\, c_{\bar{i}j}) $]
{\includegraphics[width=.33\textwidth]{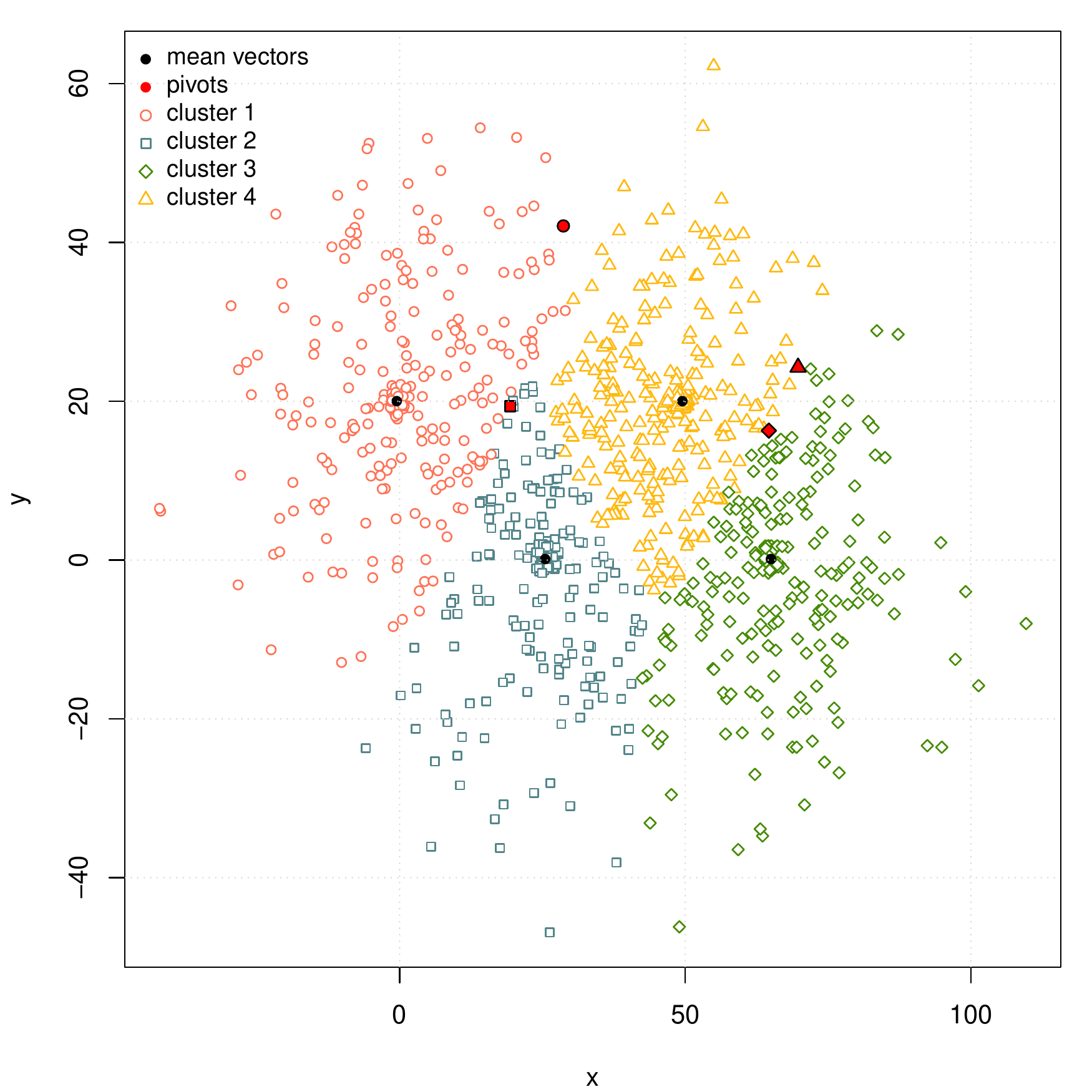}}\\
\subfloat[$\underset{\bar{i}}{\min}(\underset{j\notin\mathcal G_g}{\min}\, c_{\bar{i}j}) $] 
{\includegraphics[width=.33\textwidth]{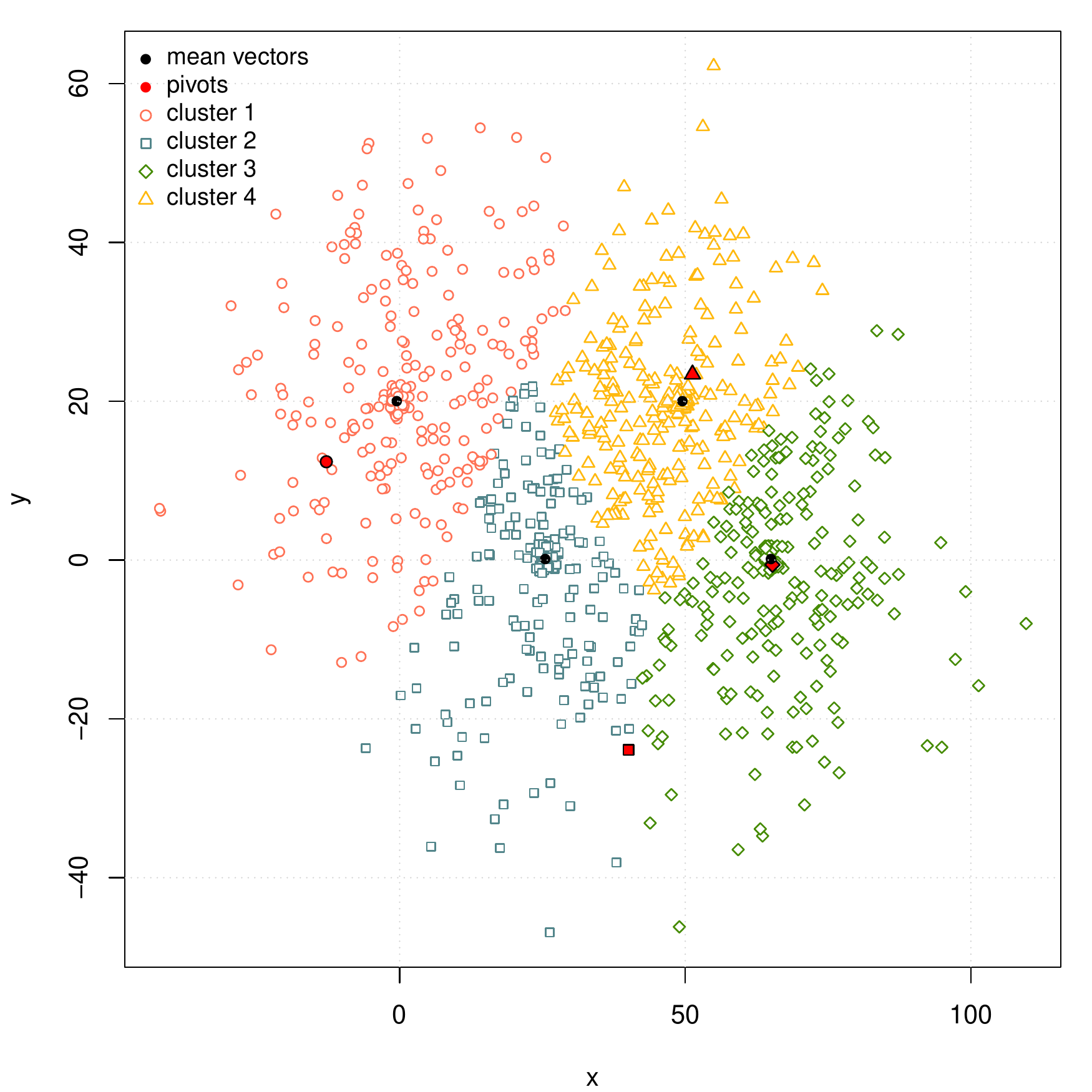}}~
\subfloat[$ \underset{\bar{i}}{\min}\sum_{j \notin \mathcal{G}_{g}} c_{\bar{i}j}$] 
{\includegraphics[width=.33\textwidth]{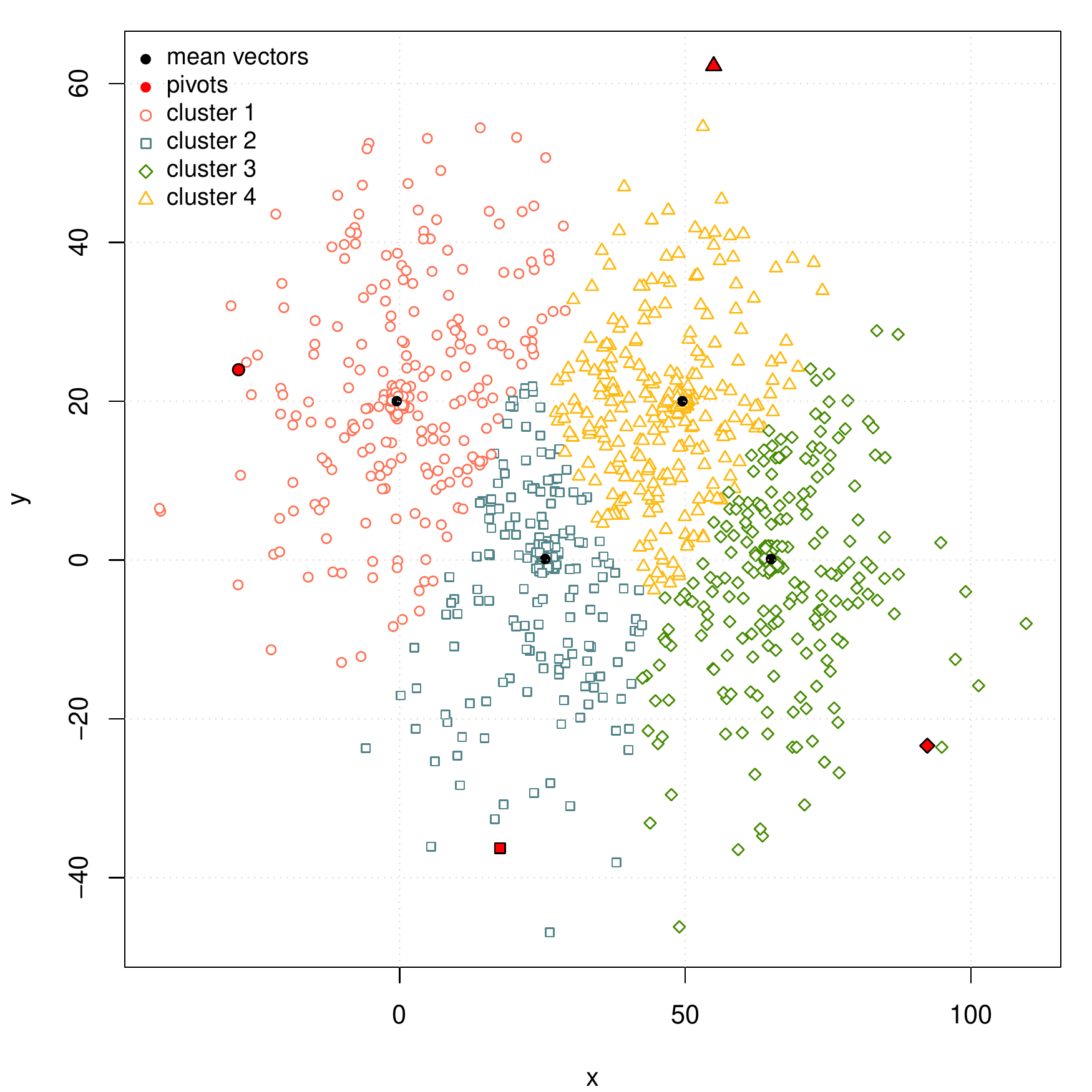}}~
\subfloat[$\underset{\bar{i}}{\max}\left(\sum_{j \in \mathcal{G}_{g}} c_{\bar{i}j} -\sum_{j \notin \mathcal{G}_{g}} c_{\bar{i}j}\right)$]
{\includegraphics[width=.33\textwidth]{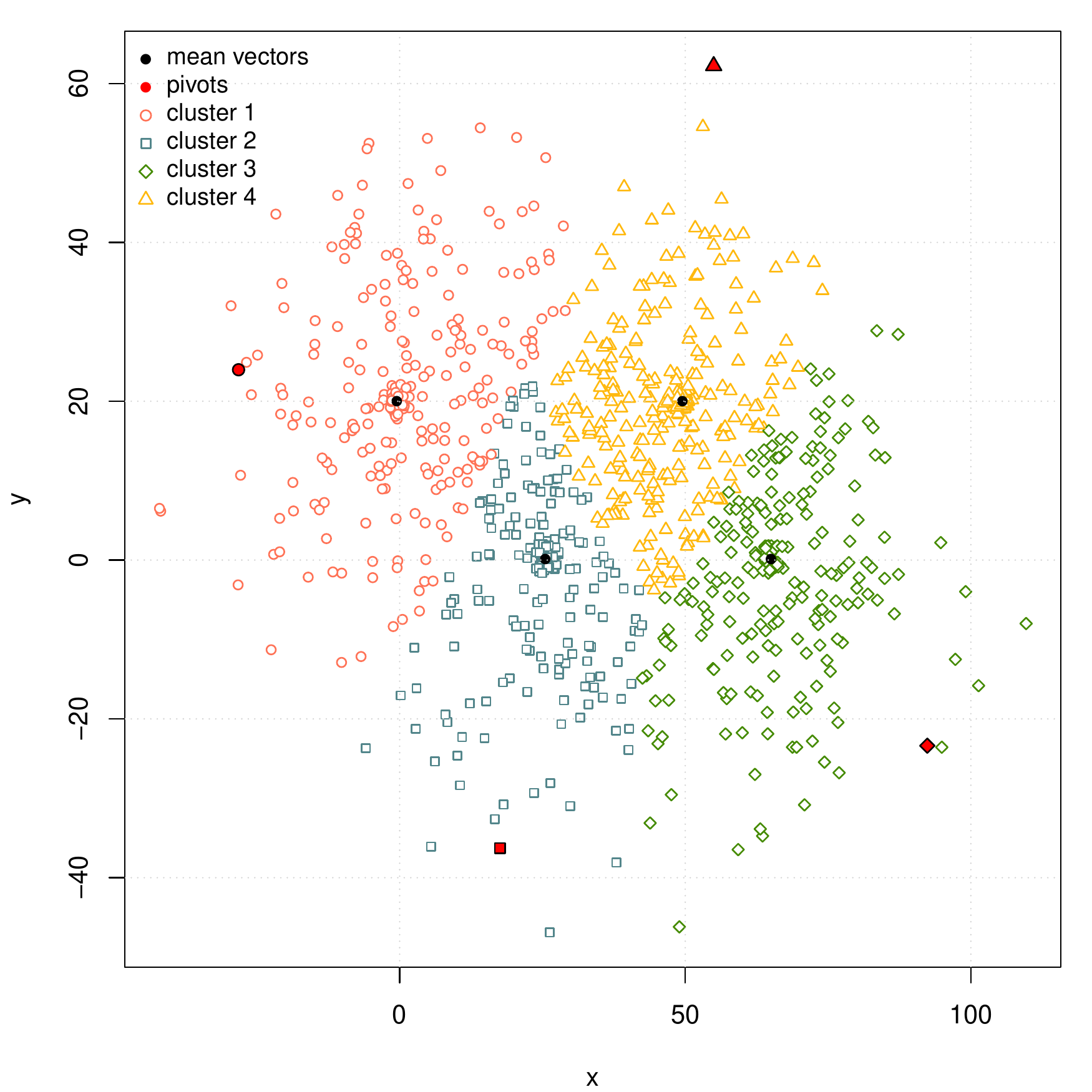}}\\
\subfloat[MUS algorithm]
{\includegraphics[width=.33\textwidth]{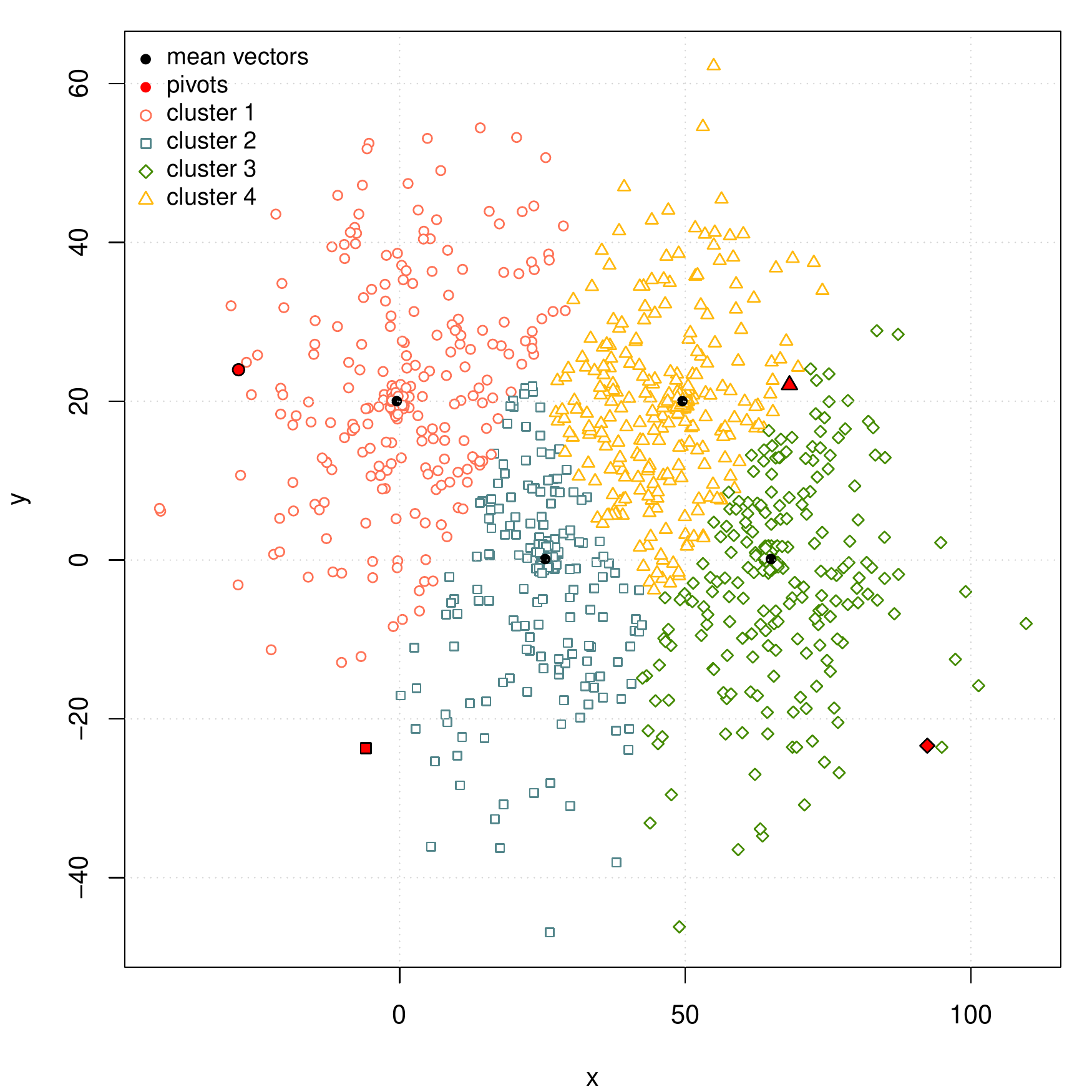}}\hfill
\vspace{-.0em} \caption{Simulated sample of size $n=1000$ from Scenario A (see Table~\ref{tab:scenarios}) clustered according to agglomerative hierarchical algorithm. The pivotal units are identified by adopting  methods (a)--(g).}
	\label{pivot:A}
\end{figure}

\begin{figure}[H]
 \centering
 \subfloat[$ \underset{\bar{i}}{\max}\left(\max_{j \in \mathcal{G}_{g}} c_{\bar{i}j}\right)$] 
{\includegraphics[width=.33\textwidth]{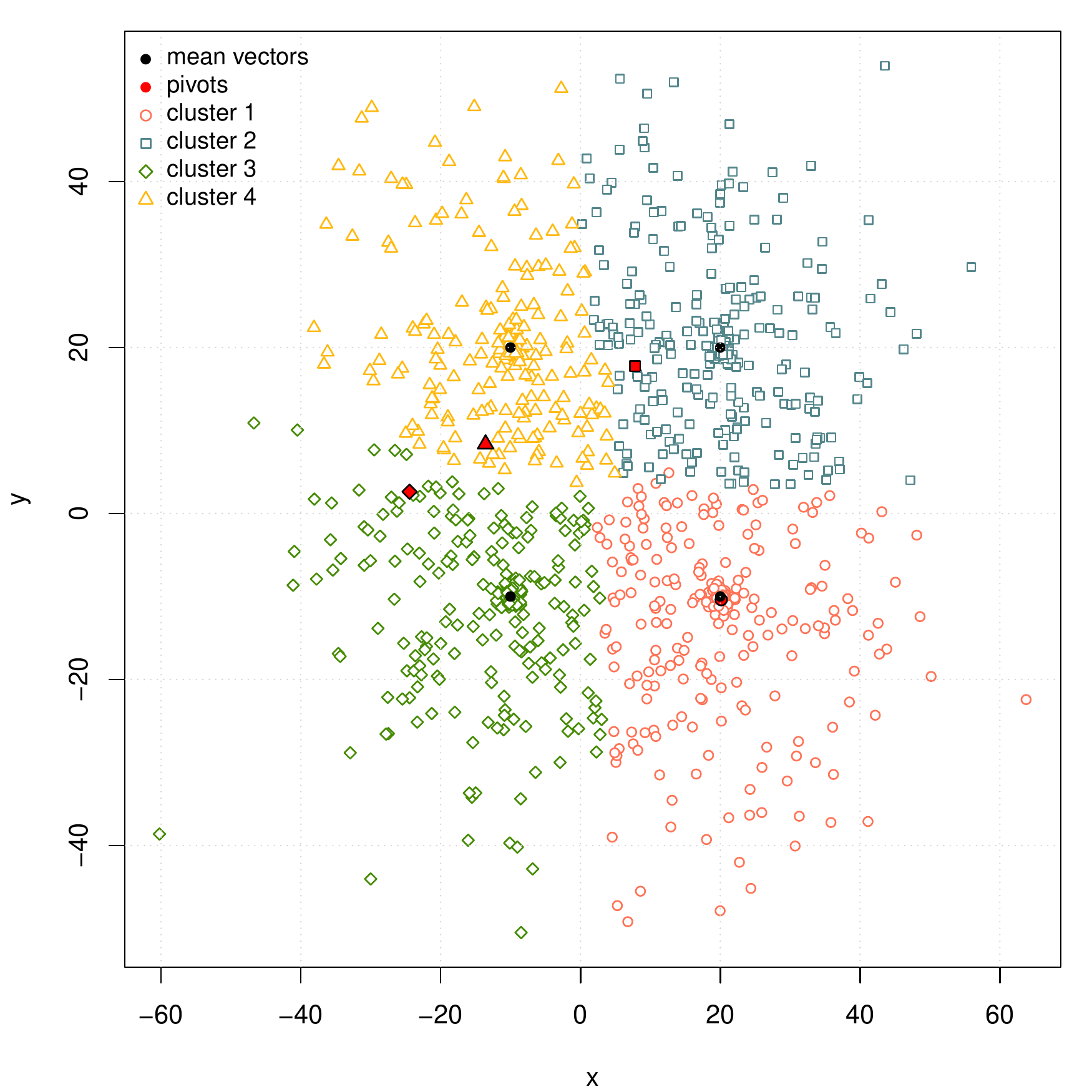}}~
\subfloat[$\underset{\bar{i}}{\max}\sum_{j \in \mathcal{G}_{g}} c_{\bar{i}j} $]
{\includegraphics[width=.33\textwidth]{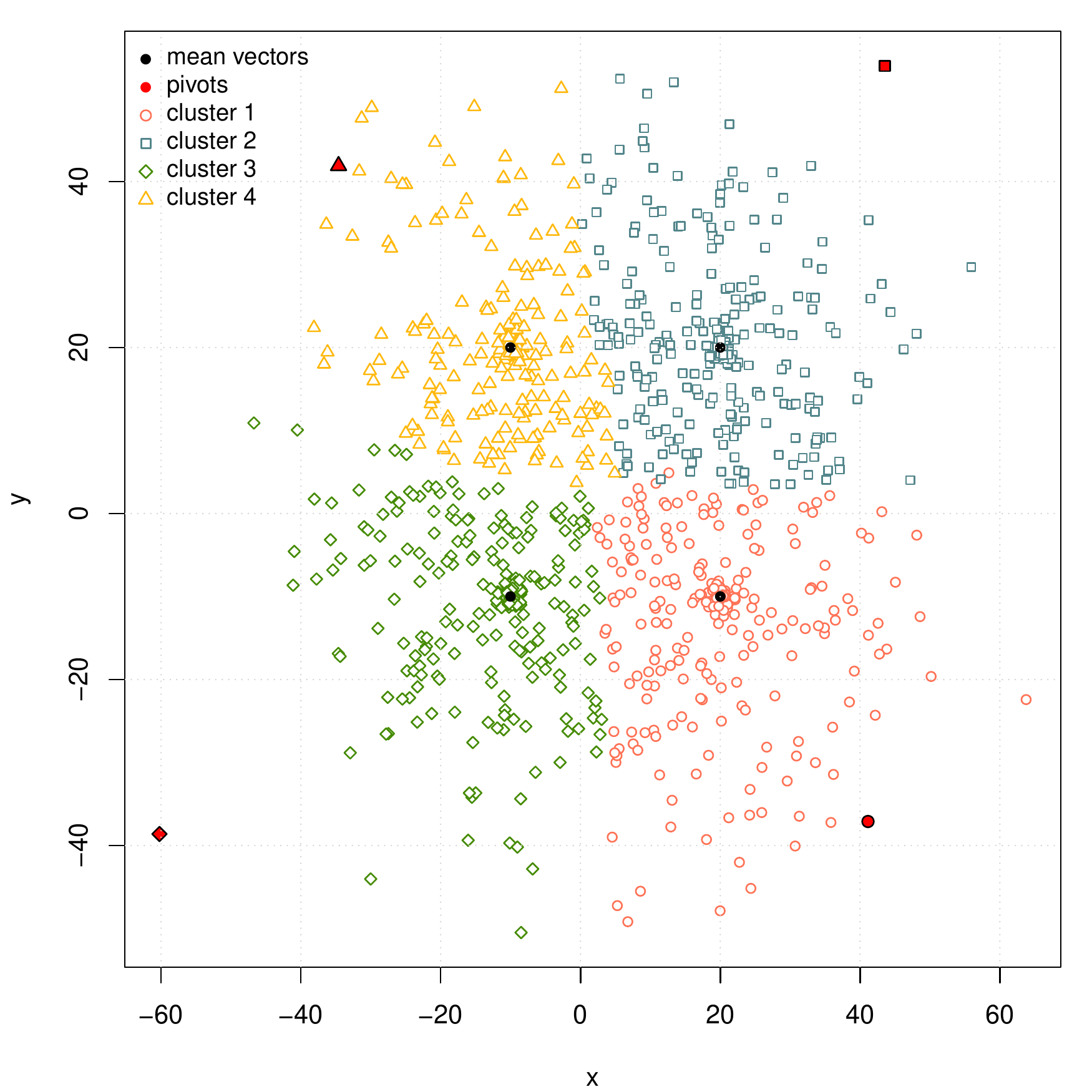}}~
 \subfloat[$\underset{\bar{i}}{\min}(\underset{j\in\mathcal G_g}{\min}\, c_{\bar{i}j}) $]
{\includegraphics[width=.33\textwidth]{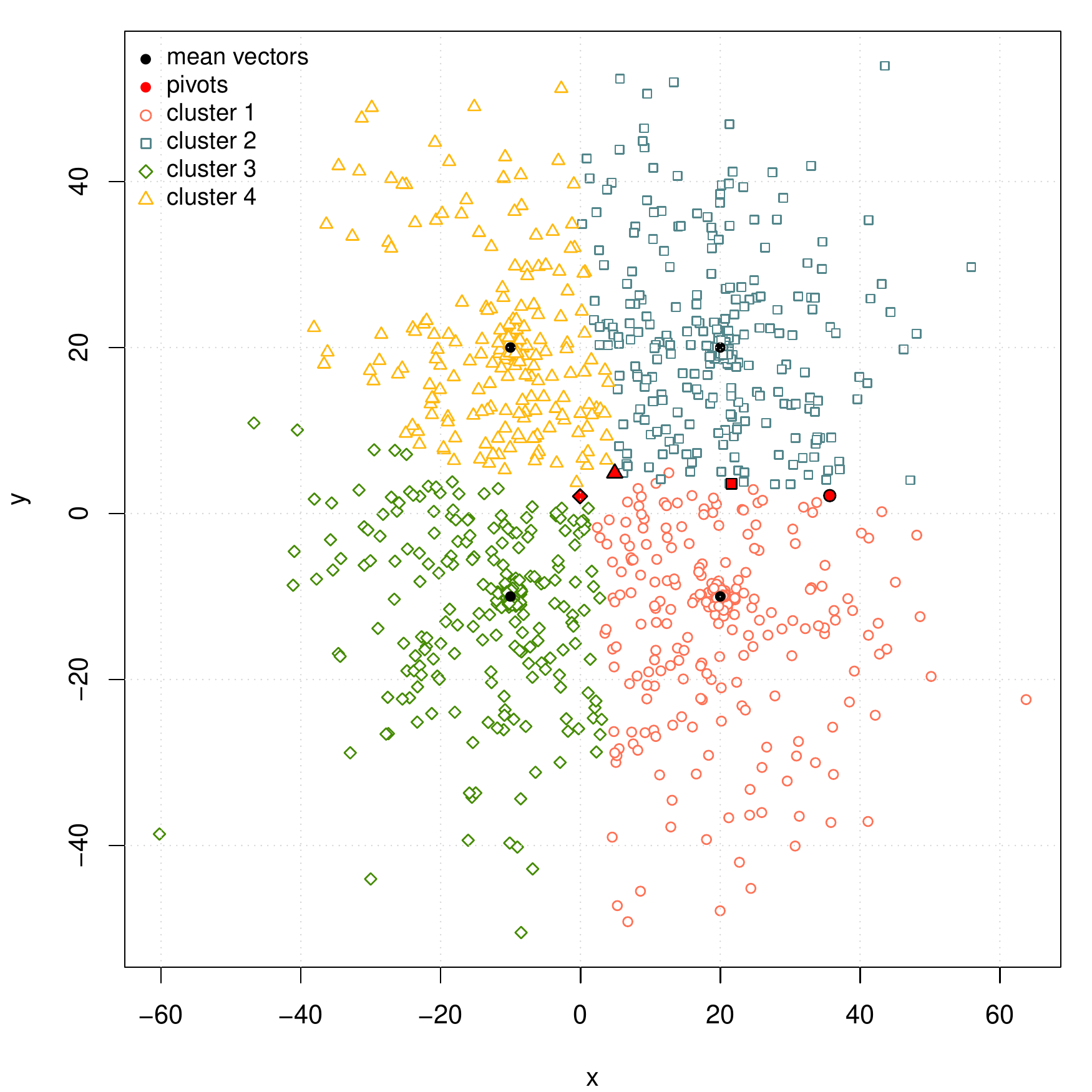}}\\
\subfloat[$\underset{\bar{i}}{\min}(\underset{j\notin\mathcal G_g}{\min}\, c_{\bar{i}j}) $] 
{\includegraphics[width=.33\textwidth]{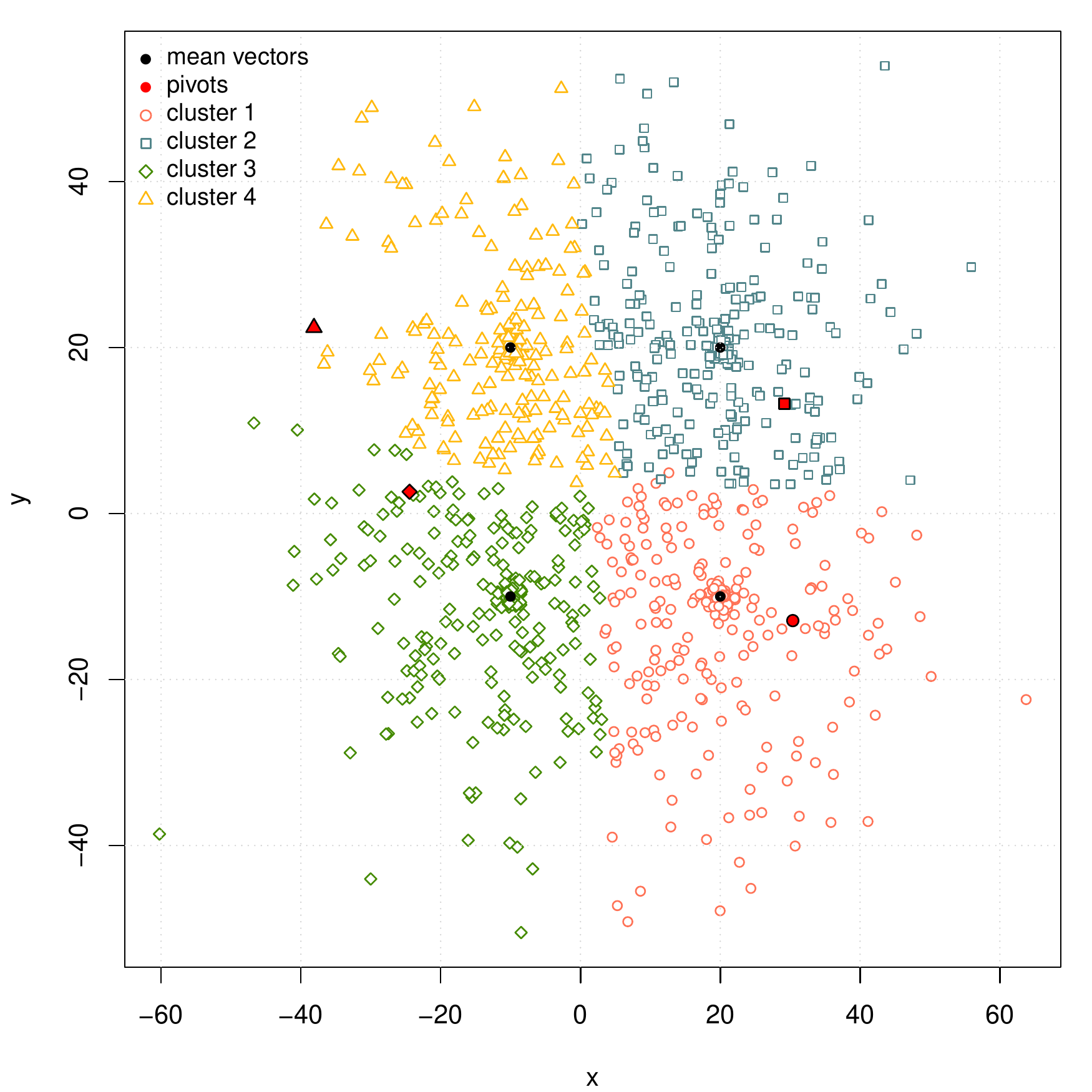}}~
\subfloat[$ \underset{\bar{i}}{\min}\sum_{j \notin \mathcal{G}_{g}} c_{\bar{i}j}$] 
{\includegraphics[width=.33\textwidth]{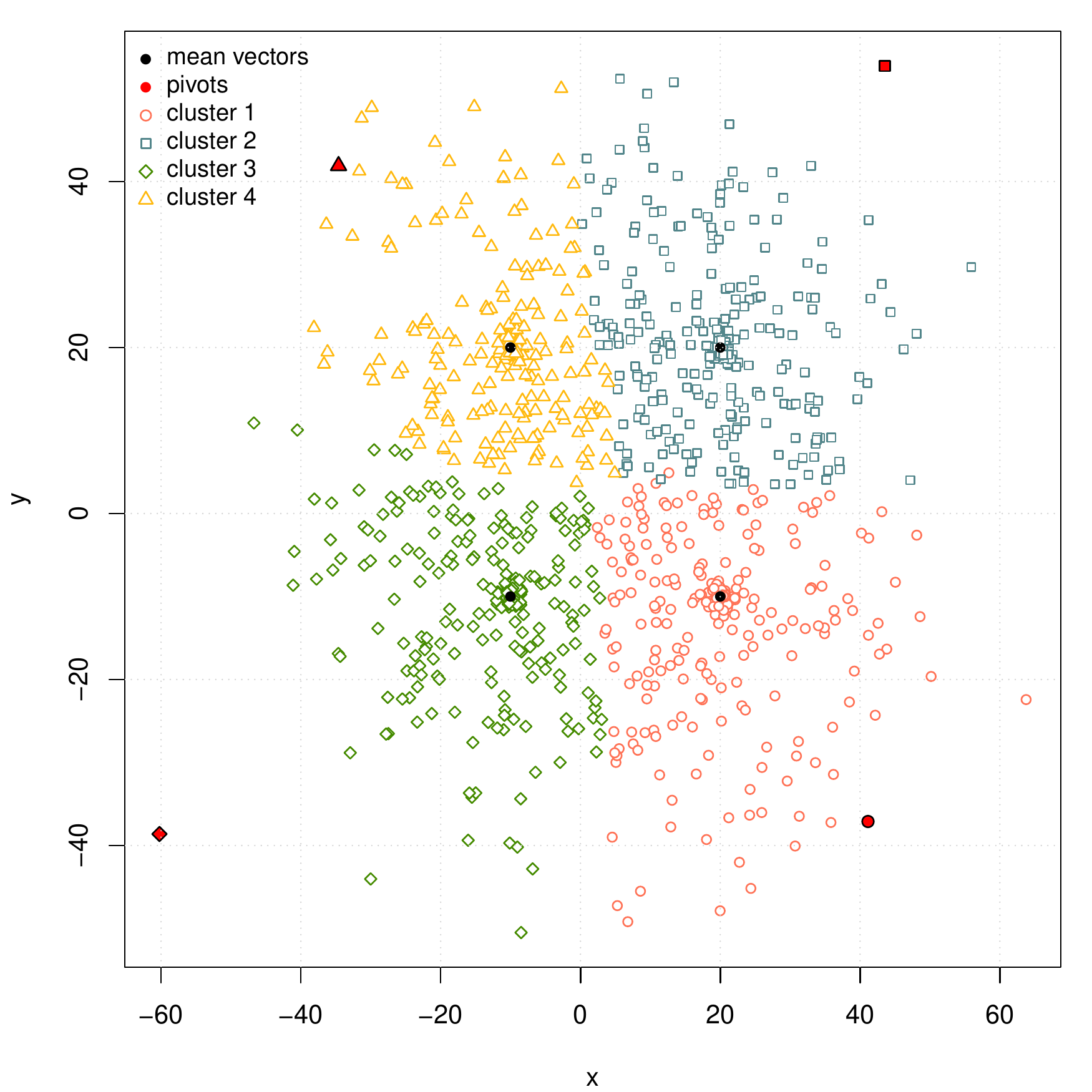}}~
\subfloat[$\underset{\bar{i}}{\max}\left(\sum_{j \in \mathcal{G}_{g}} c_{\bar{i}j} -\sum_{j \notin \mathcal{G}_{g}} c_{\bar{i}j}\right)$]
{\includegraphics[width=.33\textwidth]{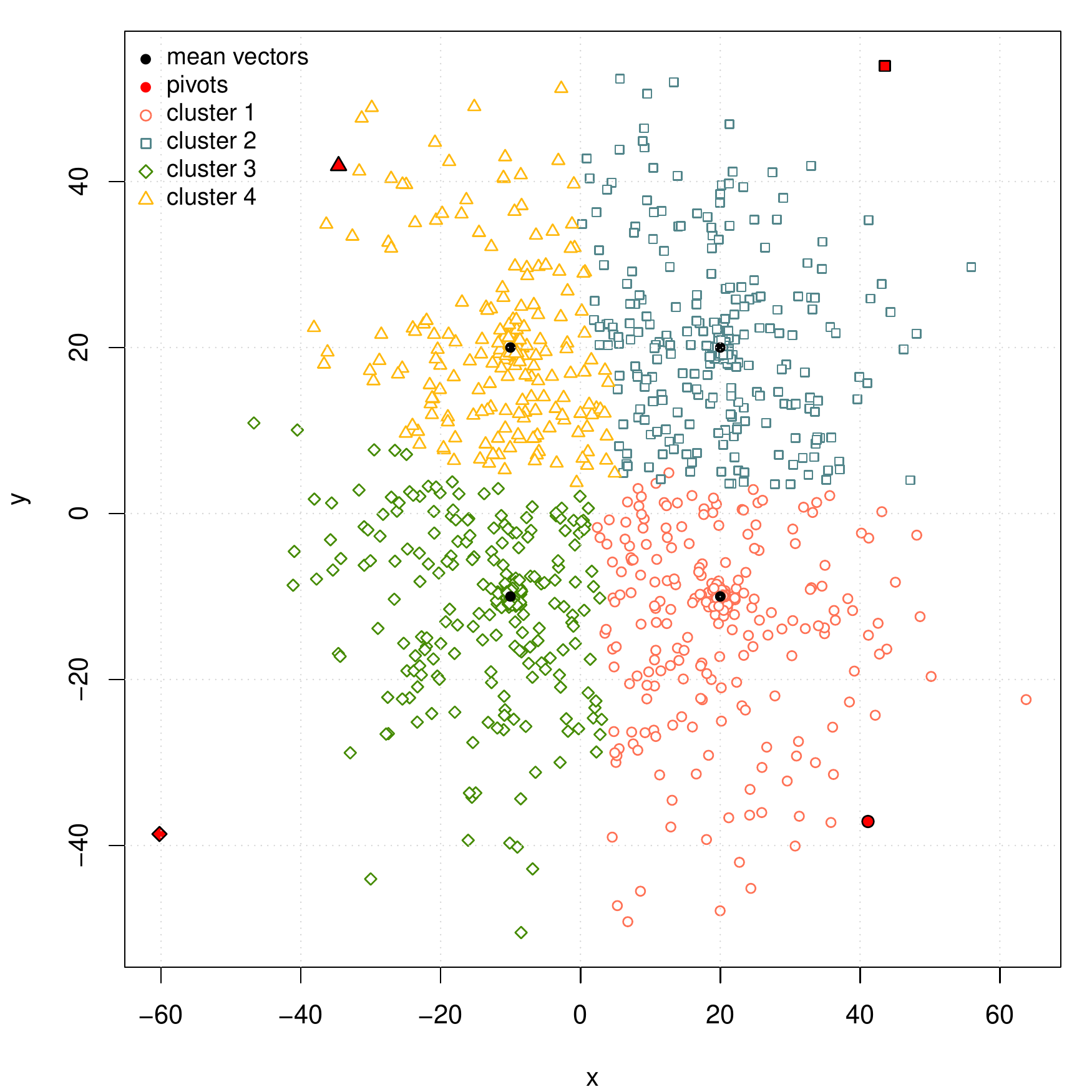}}\\
\subfloat[MUS algorithm]
{\includegraphics[width=.33\textwidth]{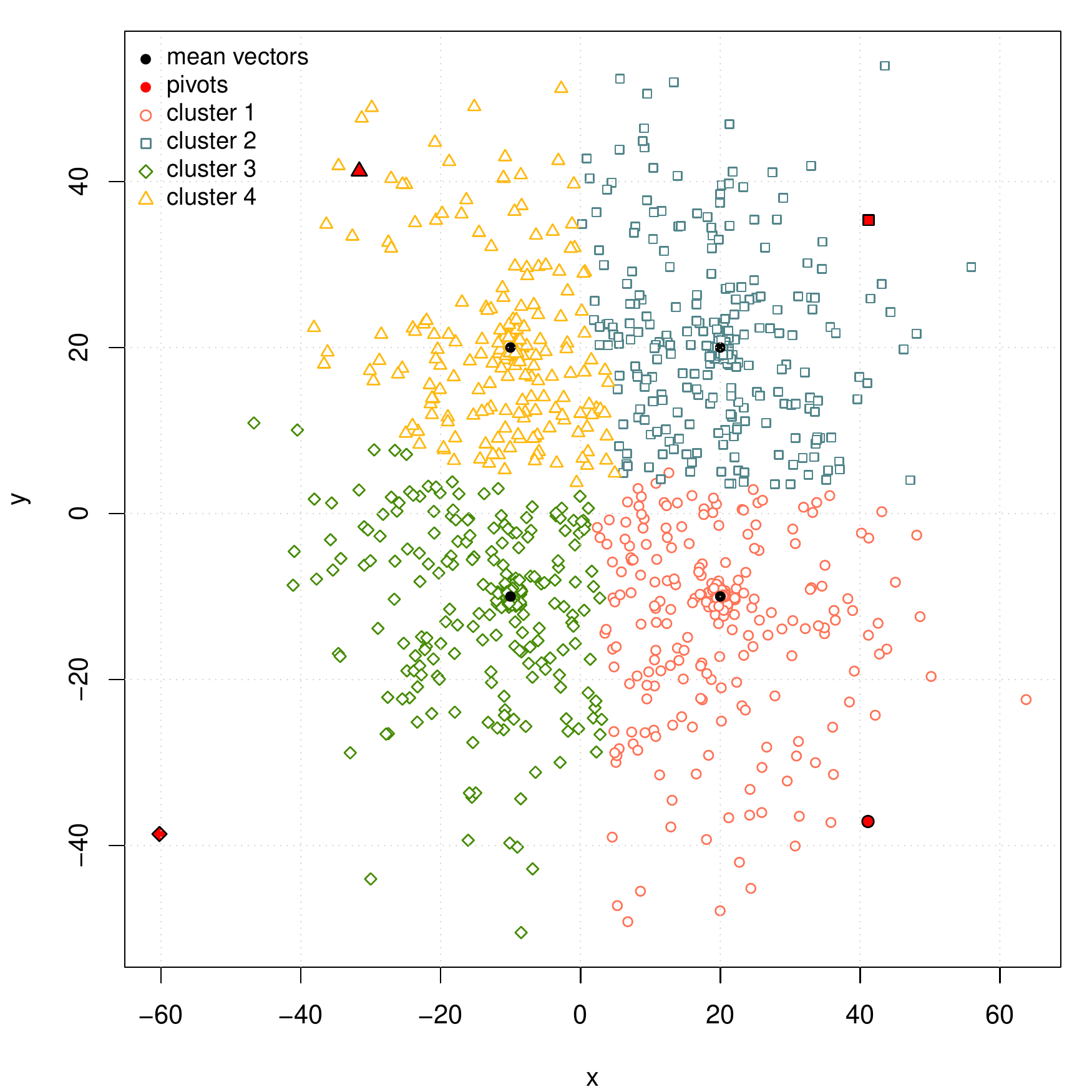}}\hfill
\vspace{-.0em} \caption{Simulated sample of size $n=1000$ from Scenario B (see Table~\ref{tab:scenarios}) clustered according to agglomerative hierarchical algorithm. The pivotal units are identified by adopting  methods (a)--(g).}
	\label{pivot:B}
\end{figure}

\begin{figure}[H]
 \centering
 \subfloat[$ \underset{\bar{i}}{\max}\left(\max_{j \in \mathcal{G}_{g}} c_{\bar{i}j}\right)$] 
{\includegraphics[width=.33\textwidth]{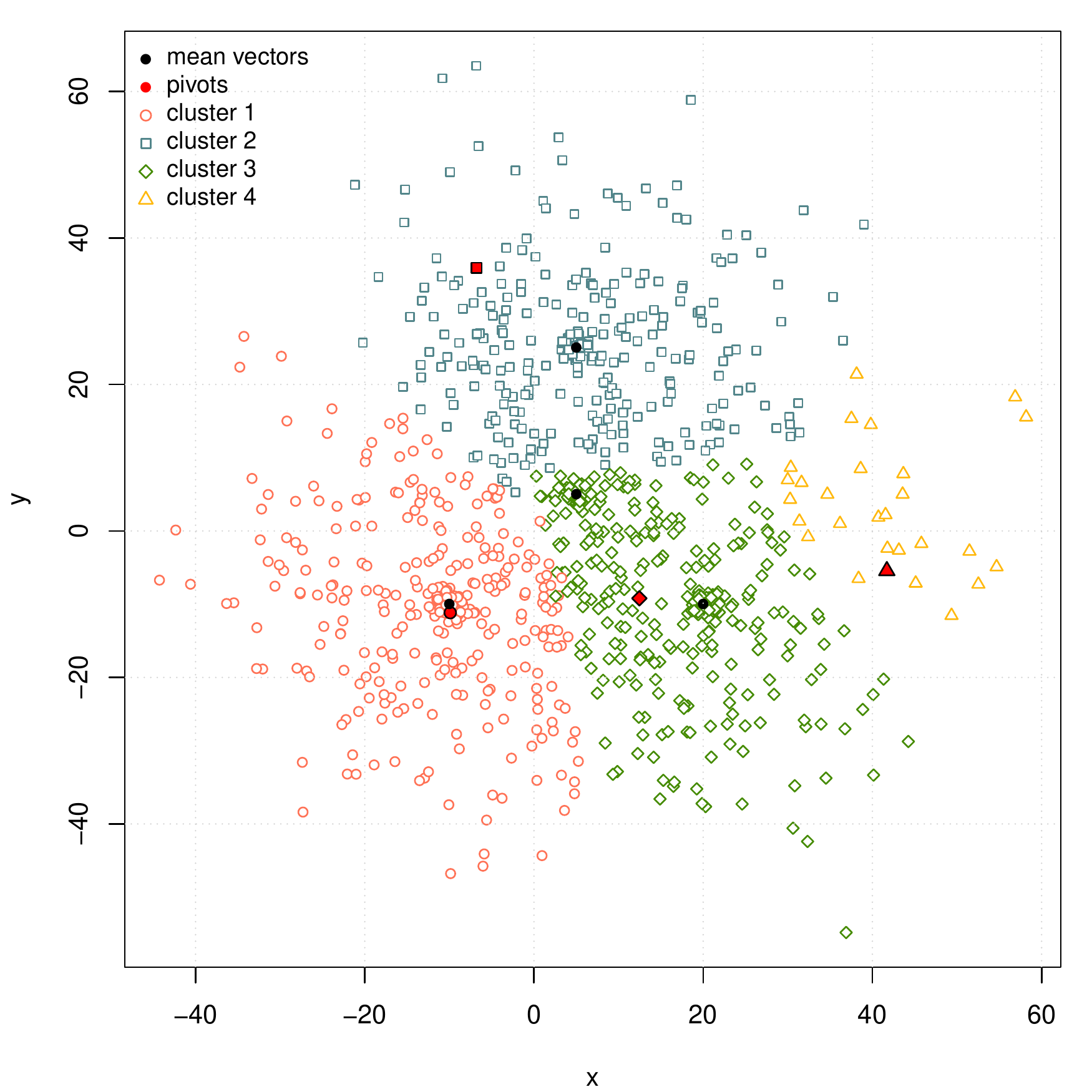}}~
\subfloat[$\underset{\bar{i}}{\max}\sum_{j \in \mathcal{G}_{g}} c_{\bar{i}j} $]
{\includegraphics[width=.33\textwidth]{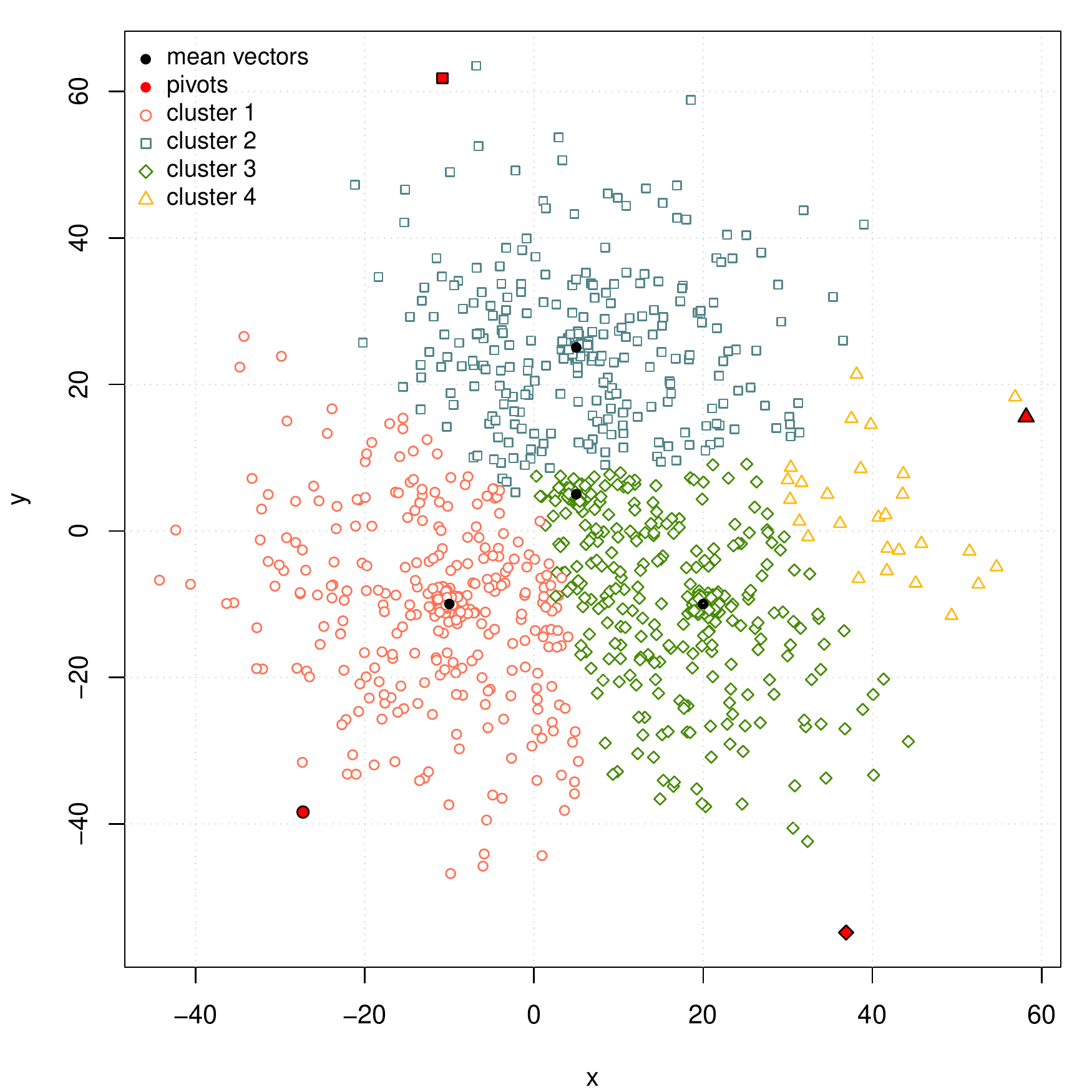}}~
 \subfloat[$\underset{\bar{i}}{\min}(\underset{j\in\mathcal G_g}{\min}\, c_{\bar{i}j}) $]
{\includegraphics[width=.33\textwidth]{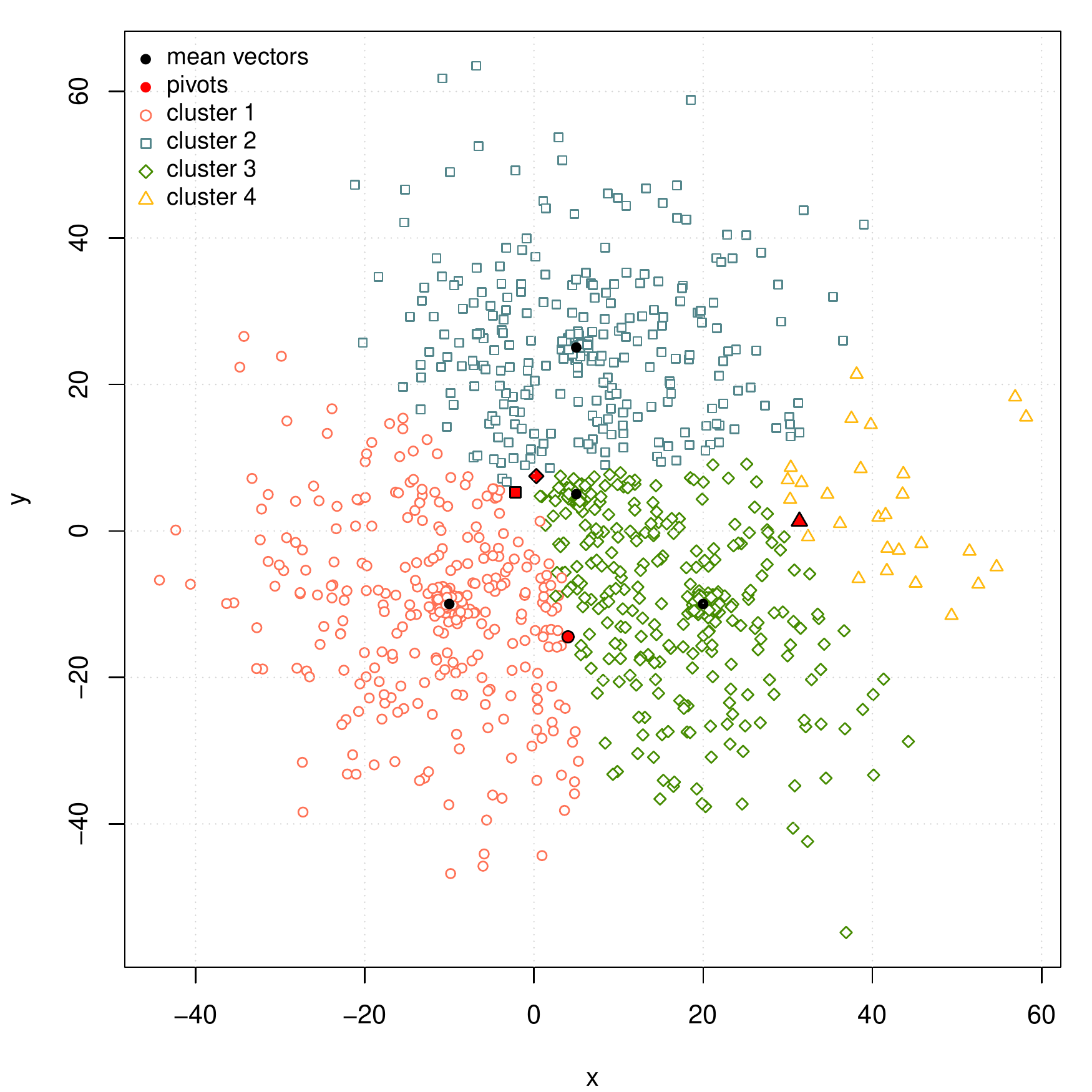}}\\
\subfloat[$\underset{\bar{i}}{\min}(\underset{j\notin\mathcal G_g}{\min}\, c_{\bar{i}j}) $] 
{\includegraphics[width=.33\textwidth]{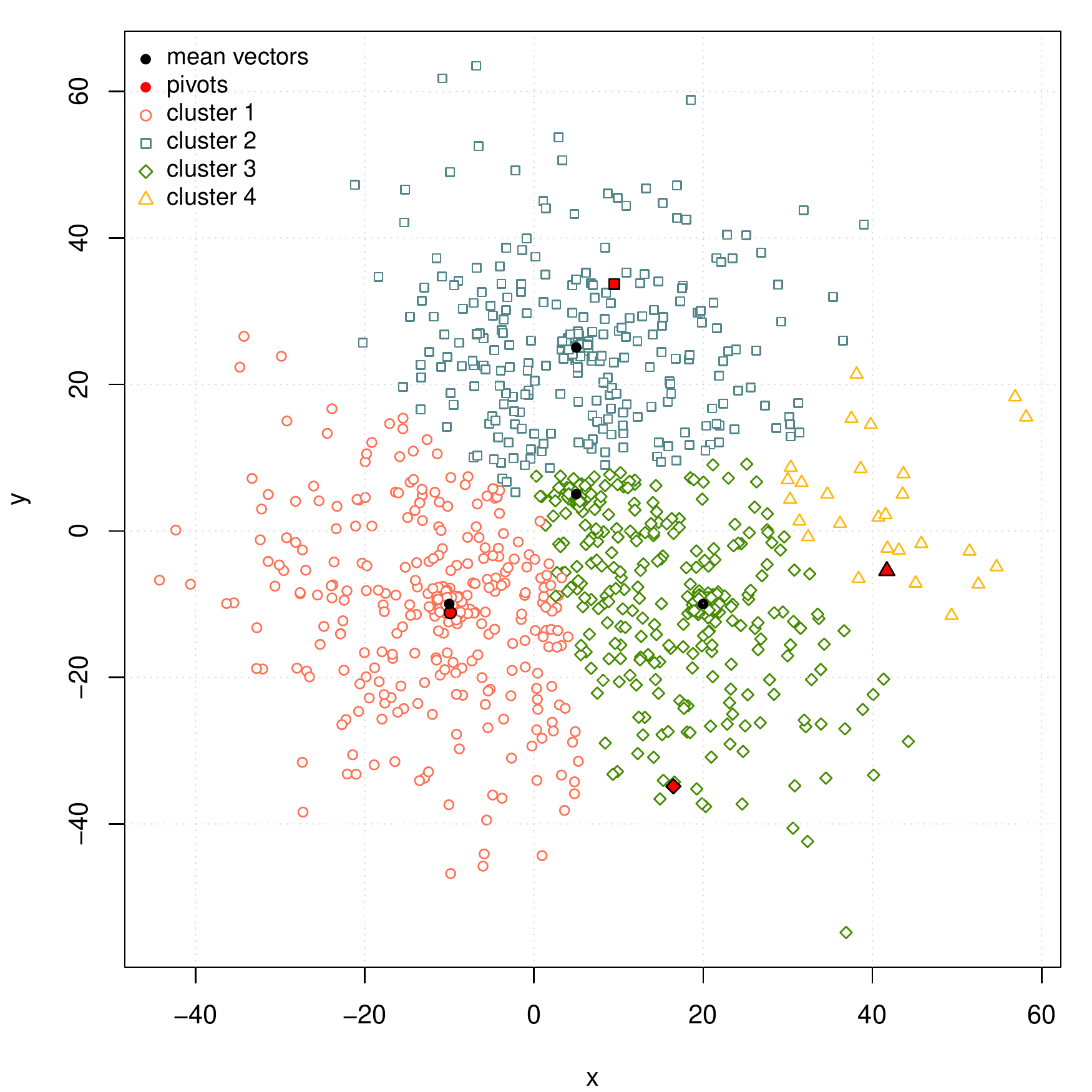}}~
\subfloat[$ \underset{\bar{i}}{\min}\sum_{j \notin \mathcal{G}_{g}} c_{\bar{i}j}$] 
{\includegraphics[width=.33\textwidth]{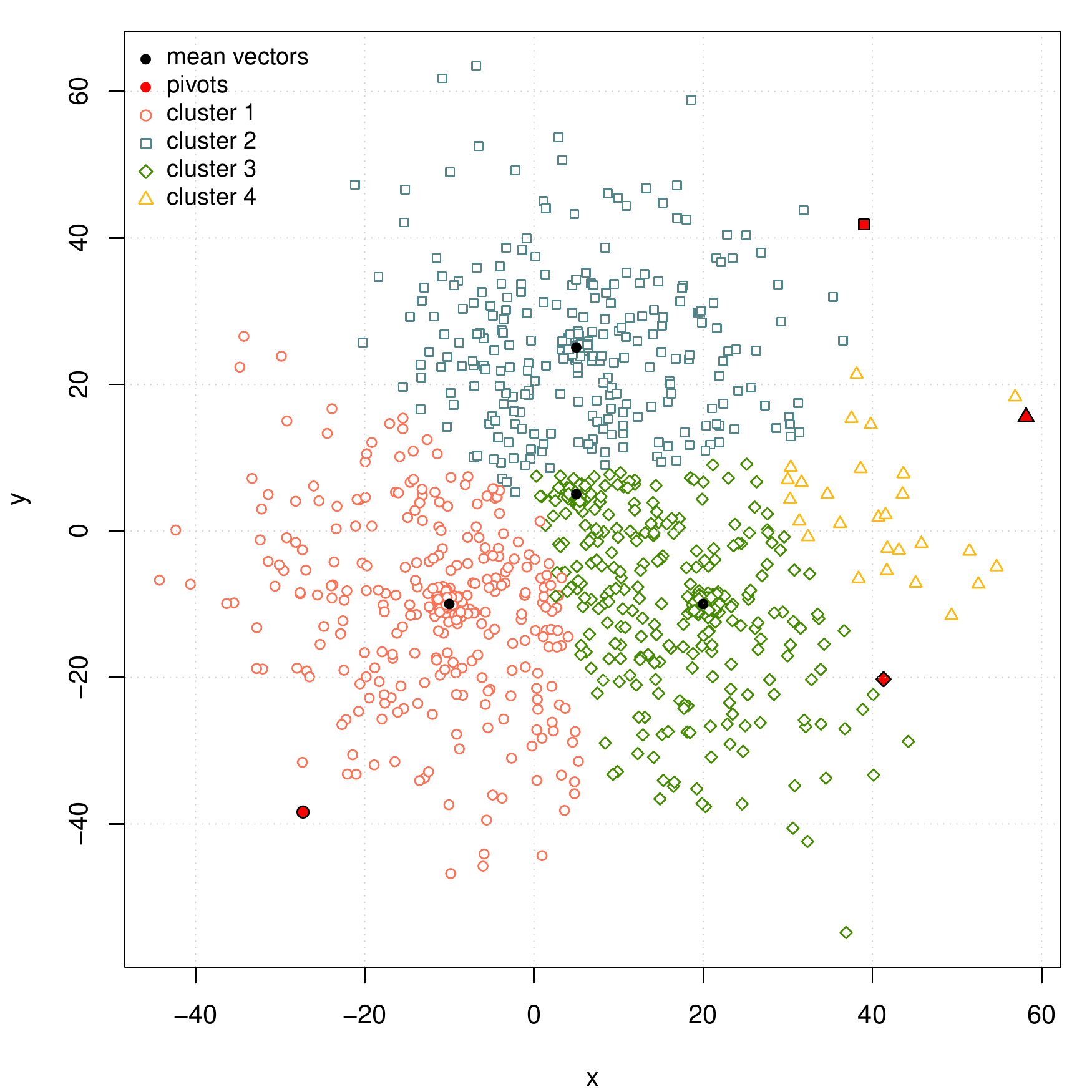}}~
\subfloat[$\underset{\bar{i}}{\max}\left(\sum_{j \in \mathcal{G}_{g}} c_{\bar{i}j} -\sum_{j \notin \mathcal{G}_{g}} c_{\bar{i}j}\right)$]
{\includegraphics[width=.33\textwidth]{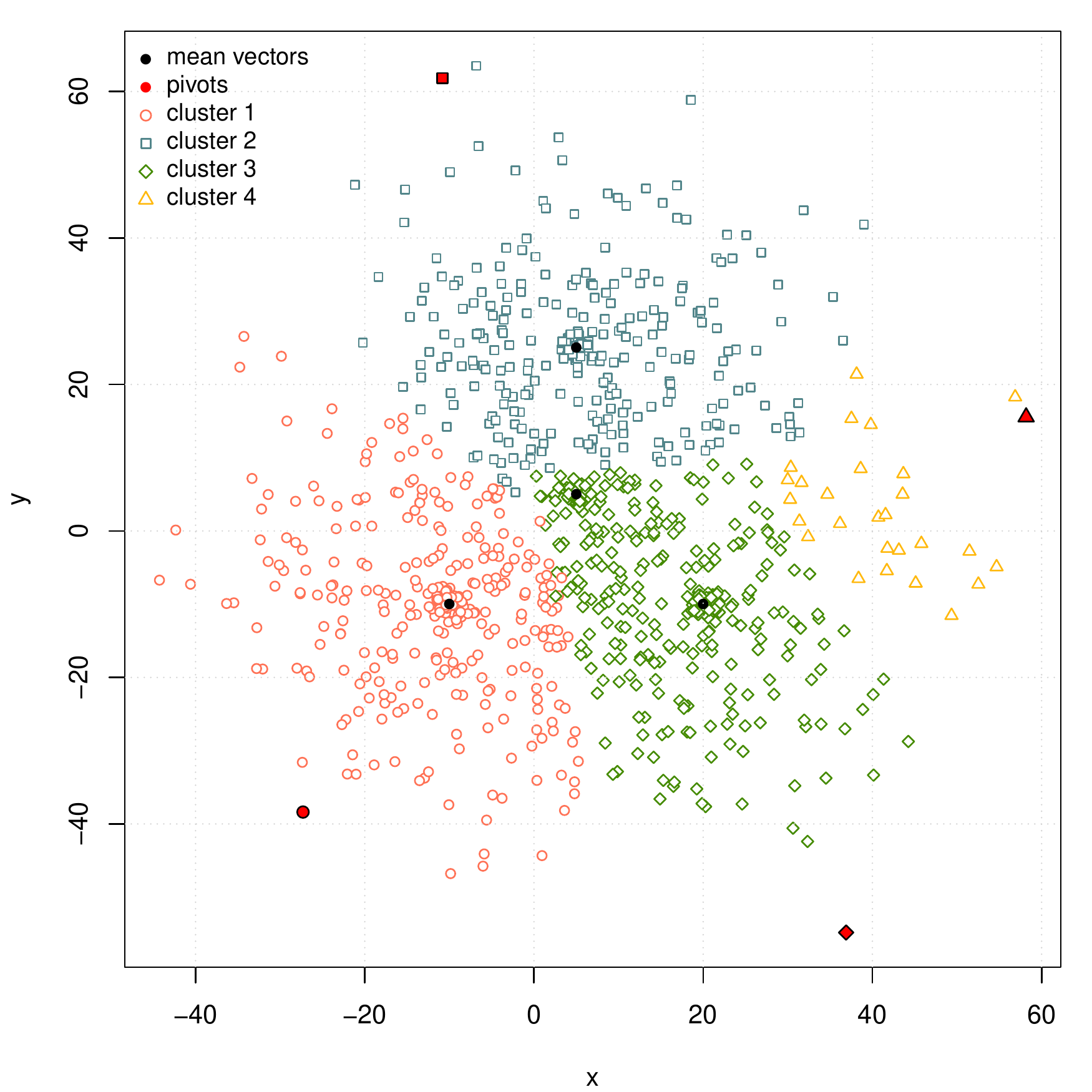}}\\
\subfloat[MUS algorithm]
{\includegraphics[width=.33\textwidth]{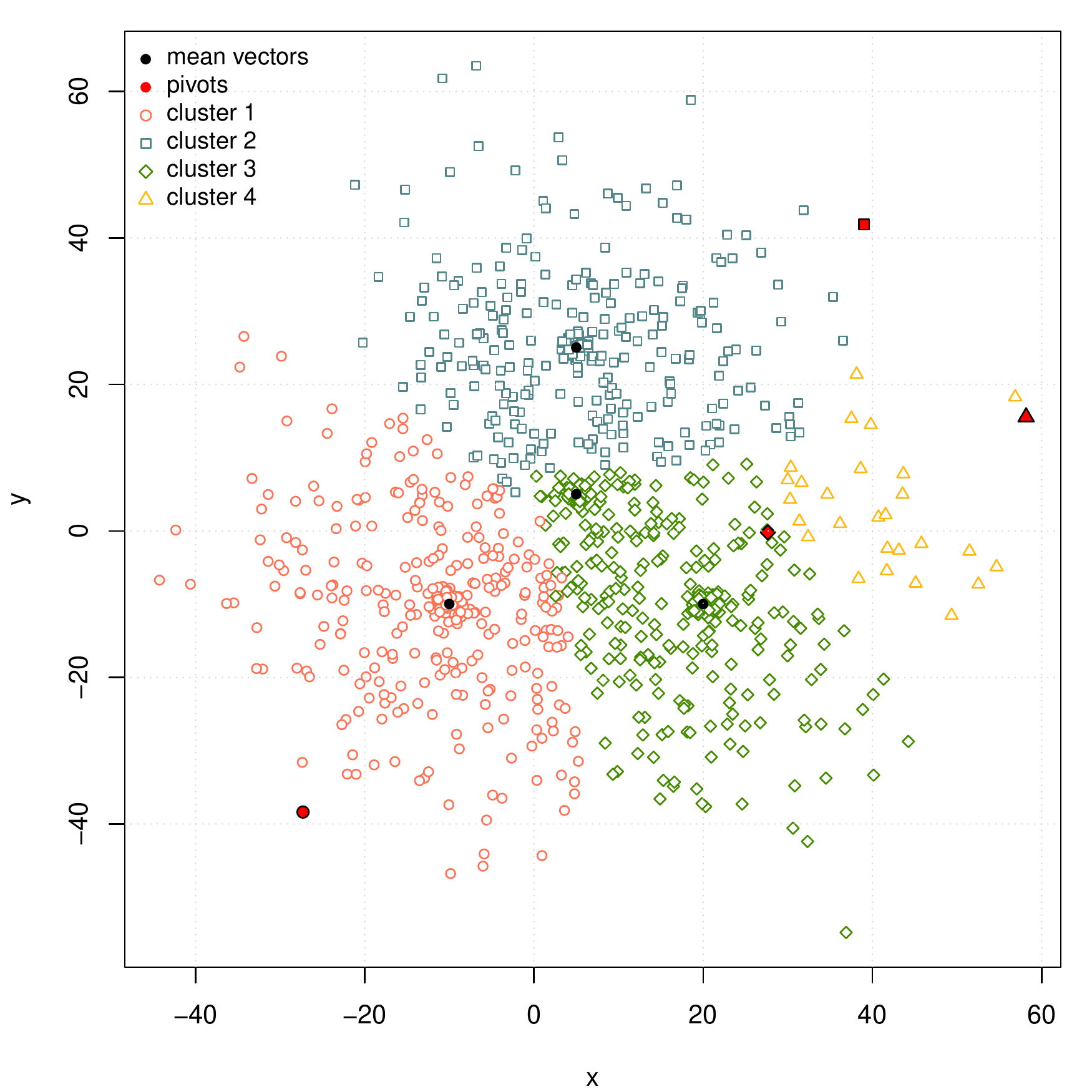}}\hfill
\vspace{-.0em} \caption{Simulated sample of size $n=1000$ from Scenario C (see Table~\ref{tab:scenarios}) clustered according to agglomerative hierarchical algorithm. The pivotal units are identified by adopting  methods (a)--(g).}
	\label{pivot:C}
\end{figure}

Figures~\ref{pivot:A}--\ref{pivot:C} display the results of the agglomerative hierarchical clustering on simulated data from scenarios A, B and C, respectively.
In each chart of Figures~\ref{pivot:A}--\ref{pivot:C} a different method for identifying the pivots is adopted ((a)-(g)), and the selected units are marked with red points on the plots. Recall that, by definition, the pivots are perfectly (pairwise) separated units. Therefore, the performance of the seven different identification methods will be higher as the posterior probability of any two of them being in the same group is closer to zero.
 As can be noticed, panels (b), (e), (f) and (g) seem to provide an accurate choice of the pivots in all situations, since they are clearly well separated and suitable as representative units for each group.
This is not a negligible issue in terms of the relabelling performance, which is strongly affected by the choice of such $\hat G$, by virtue of Equation~\eqref{eq:relabelmu}. The estimated proportions of relabelled iterations based on 100 simulated samples are reported in Table~\ref{tab:prop}. As expected, better performances in terms of pivot selection are likely to reflect into a higher proportion of relabelled iterations in almost all situations. 

Coherently with the considerations drawn from Figures~\ref{pivot:A}--\ref{pivot:C}, methods (b), (e) and (f) register the highest chain proportions (less than $1\%$ of the iterations is discarded) for both scenarios A and B. Method (c) seems to have the worst performance regardless of the considered scenario, in particular, for scenario C the chain keeps only about $8\%$ of the original iterations. The fact that the third simulated scenario shows globally less satisfactory results is not surprising. In fact, the input means are so close each other that the cluster algorithms may fail in recognizing the true data partition, thus reflecting on the quality of the choice of the pivotal units. However, (e) and (f) criteria and MUS algorithm are preferable to the others.

\begin{figure}[H]
\centering
\includegraphics[width=\textwidth]{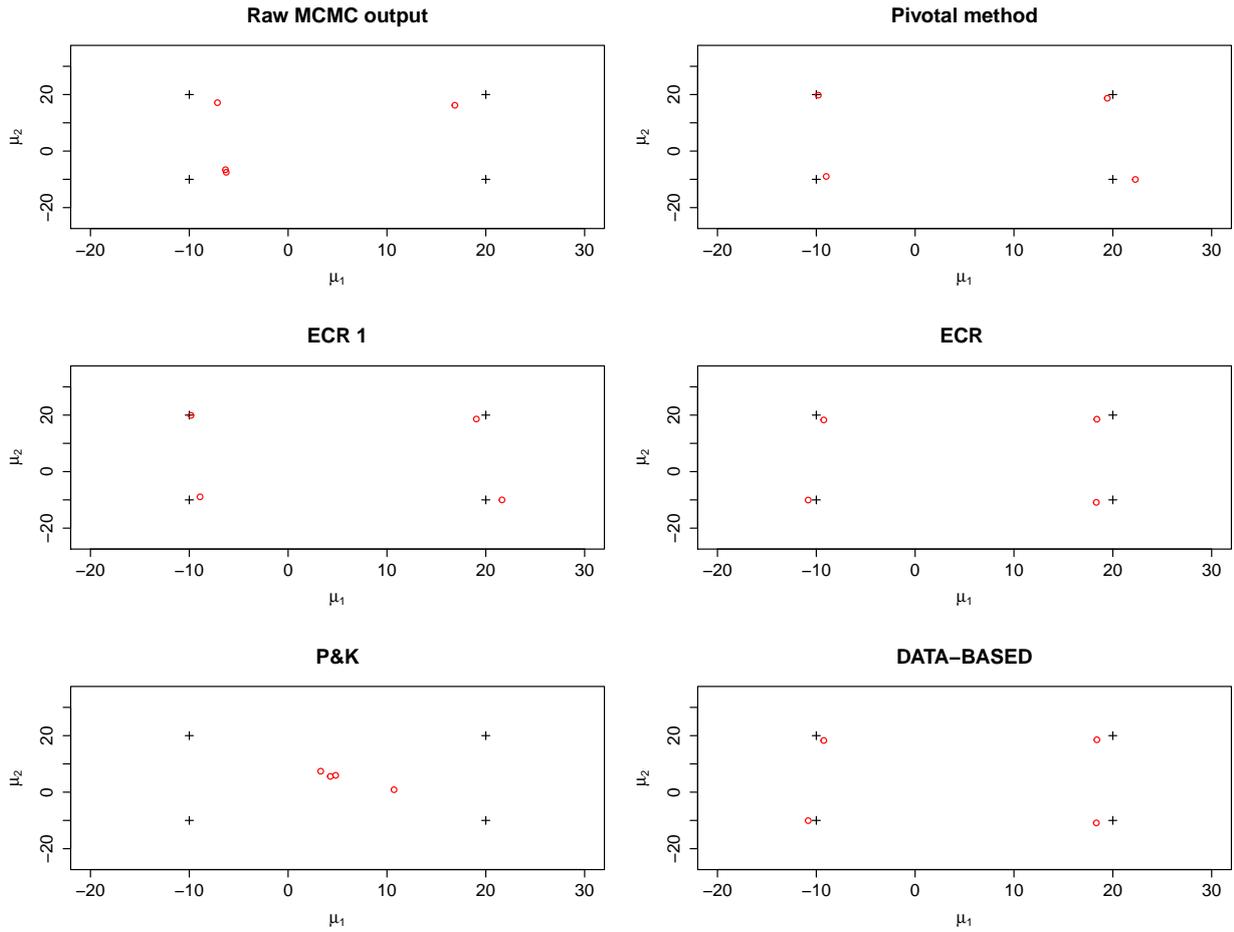}
\caption{Scenario B. Crosses are input means, red points
are the median values of relabelled estimates.  \emph{(Top left)} Raw MCMC sample for $\mu_{g}, g=1,...,4$. \emph{(Top right)} Reordered MCMC sample of Pivotal Method resulting from agglomerative hierarchical clustering and MUS algorithm.
Reordered MCMC sample
according to methods ECR-iterative-1, ECR, Puolam\"{a}ki and Kaski and DATA-BASED.}
\label{6metodi} 
\end{figure}

In order to compare the proposed methodology with other relabelling algorithms, in the task of estimating the means of the mixture components, we consider the Puolam\"{a}ki and Kaski procedure \citep{puolamaki2009bayesian} and three other methods implemented in the \textbf{label.switching} package \citep{papastamoulis2016label}. 
In Figure~\ref{6metodi} the median estimates of relabelled group means are plotted for a simulated example from scenario B and four alternative methods. As can be seen, our relabelling procedure seems to provide very accurate estimates of group means. Similar results are achieved by ECR-iterative-1, ECR and DATA-BASED, while  Puolam\"{a}ki and Kaski algorithm appears not to yield reliable estimates for the group means.

In Table~\ref{tab:MSE} are reported the mean square errors of the relabelled estimates, obtained as mean over $B=100$ macro-replications of the Euclidean distances between the input means and the corresponding estimates, according to scenarios A, B and C. 
In all scenarios the highest mean square errors are obtained for method (c), for each component of the mixture. Criteria (b), (e) and (f) give very similar results, and the MUS algorithm outperforms all other pivotal methods for three cases in Scenario A and two in Scenario B. ECR-iterative-1 performances in terms of mean square errors are comparable with our algorithm in most cases, although it shows to give lower errors in the third scenario, while Puolam\"{a}ki and Kaski algorithm seems to provide poor estimates in all situations (see also Figure~\ref{6metodi}).

\begin{table}[H]
	\centering
	\begin{tabular}{rrrrrrrr}
	\hline
		 & (a) &  (b) &  (c) &  (d) &  (e) &  (f) &  (g) \\
		\hline
		Scenario A & 0.475 & 0.993& 0.124 & 0.506 & 0.993 & 0.993 & 0.313 \\
		Scenario B & 0.519 & 0.998 & 0.101 & 0.707 & 0.998 & 0.998 & 0.995 \\ 
		Scenario C & 0.139 &   0.300 &  0.079 &  0.267 &  0.368 &   0.507 &   0.374 \\
		\hline
	\end{tabular}
	\medskip
	\caption{Estimated proportion of relabelled iterations (see  Equation~\eqref{eq:condevent}), over 100 macro-replications, based on the original MCMC sample, according to Scenario A, B and C. The observations are clustered according to agglomerative hierarchical algorithm.} 
	\label{tab:prop}
\end{table}

\newpage
\newpage
\newpage

\begin{table}[H]
\centering
\begin{tabular}{rrrrrr}
\hline
& Scenario A & $\bm{\mu}_{1}$ & $\bm{\mu}_{2}$ & $\bm{\mu}_{3}$ & $\bm{\mu}_{4}$ \\ 
  \hline
 & (a) & 13.8078 & 1.6724 & 2.0158 & 10.8232  \\ 
  &(b) & 13.7064 & 1.6104 & 1.9814 & 9.1846   \\ 
  &(c) & 22.8786 & 7.5307 & 6.7326 & 14.5896 \\ 
  &(d)  & 14.0215 & 1.6619 & 1.9951 & 11.2910 \\ 
  &(e)  & 13.7301 & 1.6264 & 1.8889 & 9.2900 \\ 
  &(f) & 13.7794 & 1.6723 & 1.8979 & 9.2897 \\ 
  & MUS  & \textbf{12.5787} & \textbf{1.5531} & \textbf{1.7919} & 9.6220  \\ 
   \hline
   &ECR-1 & 13.6403 &  1.6605 &  1.9015 & \textbf{8.8085}  \\
   &P \& K & 25.5940 & 15.5229 & 15.1522 & 27.2411 \\\hline
   \hline
& Scenario B & $\bm{\mu}_{1}$ & $\bm{\mu}_{2}$ & $\bm{\mu}_{3}$ & $\bm{\mu}_{4}$ \\
  \hline
 & (a) & 1.4066 & 1.6251 & 1.5984 & 1.5884 \\ 
  &(b) & 1.4123 & 1.6005 & 1.5737 & 1.5419 \\ 
  &(c) & 4.8496 & 4.3588 & 4.8097 & 5.2142 \\ 
  &(d) & 1.4096 & 1.5961 & 1.5729 & 1.5403 \\ 
  &(e) & 1.4127 & 1.6003 & 1.5736 & 1.5417 \\ 
  &(f) & 1.4121 & 1.5982 & 1.6192 & 1.5420 \\ 
  & MUS  & \textbf{1.4070} & \textbf{1.5877} & 1.5728 & 1.5437 \\ 
   \hline
   &ECR-1 & 1.4129 & 1.5984 & \textbf{1.5717} & \textbf{1.5429} \\
   &P \& K & 18.4657 & 18.6185 & 18.6796 & 19.0404\\\hline
   \hline
& Scenario C & $\bm{\mu}_{1}$ & $\bm{\mu}_{2}$ & $\bm{\mu}_{3}$ & $\bm{\mu}_{4}$ \\
  \hline
  & (a) & 9.1013 & 9.9974 & 8.4766 & 19.3288 \\  
  & (b) & 6.9196 & 7.8994 & 8.7700 & 14.1766 \\ 
  & (c) & 11.6894 & 10.8810 & 8.8252 & 22.6435 \\ 
  & (d) & 7.7730 & 9.1701 & 9.1987 & 16.4153 \\
  & (e)  & 7.6160 & 7.1054 & 10.2073 & 13.2589 \\ 
  & (f) & 7.1992 & 7.1643 & 9.4728 & 15.2713 \\ 
  & MUS & 6.7458 & 7.5579 & 9.7924 & 14.8356 \\ 
   \hline
   &ECR-1 & \textbf{6.4891} & \textbf{6.7234} & 8.4472 & \textbf{9.3649}\\
   &P \& K & 17.5726 & 16.8717 &  \textbf{3.4988} & 20.2620\\
     \hline
\end{tabular}
\caption{Mean squared error $\displaystyle\dfrac{1}{B}{\sum_{j=1}^{B}}||\vmu_{gs}^{(j)}-\hat{\vmu}_{gs}^{(j)}||$ computed for $B=100$ macro-replications, of the estimates of the mean vector components $\vmu_{gs}$,  $g=1,...,4, s=1,2$, according to criteria (a)--(f) and MUS of pivotal method, Puolam\"{a}ki and Kaski (P\&K) and ECR-1 algorithms.}
\label{tab:MSE}
\end{table}
\newpage

\section{Case study}
\label{case:study}

\begin{figure}[H]
\begin{center}
\includegraphics[scale=0.5]{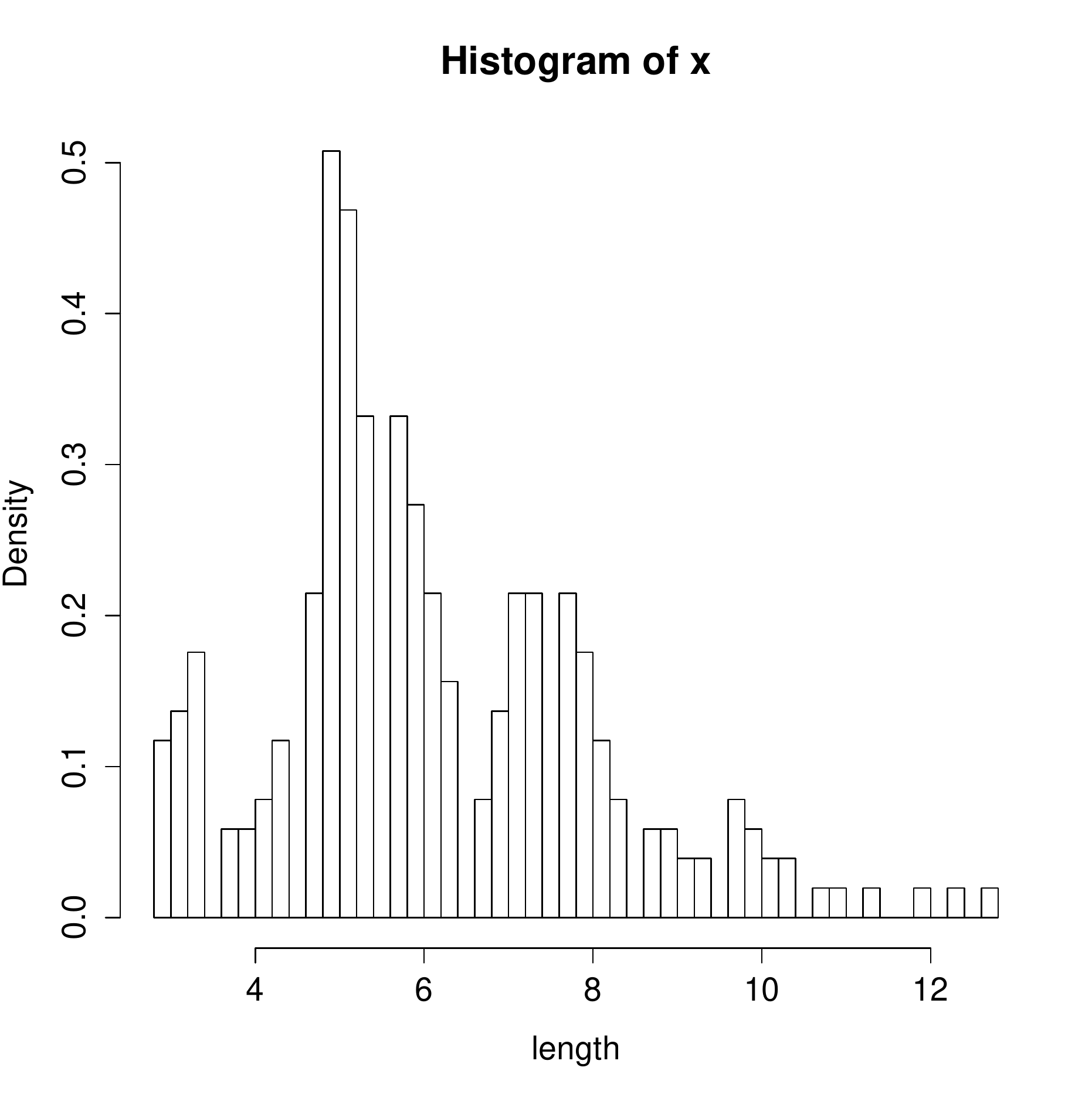}
\caption{Histogram of fishery data. On $x$-axis the snapper length measurements}
\label{fish}
\end{center}
\end{figure}

Fishery dataset, originally taken from \citet{titterington1985statistical} and used by \citet{papastamoulis2016label} for comparing different relabelling procedures, consists of $n = 256$ snapper length measurements. In Figure~\ref{fish} the histogram of the lengths is shown. 
In this section, we apply our method to these data and test its efficiency, comparing the results with some methods from the \textbf{label.switching} package. 
We use a Gaussian mixture with $G=5$ components as suggested by \citet{papastamoulis2016label}, that is:

\begin{equation}
\label{eq:fish}
y_{i} \sim \sum_{g=1}^{G} p_{g} \mathcal{N}(\mu_{g}, \sigma^{2}_{g}), \ \ i=1,...,n
\end{equation} 

We set up a Gibbs sampling through the \textbf{bayesmix} \texttt{R} package \citep{grun2011bayesmix}, with $H=11000$ iterations and a burn-in period of 1000. 

\begin{figure}[H]
\begin{center}
\includegraphics[scale=0.5]{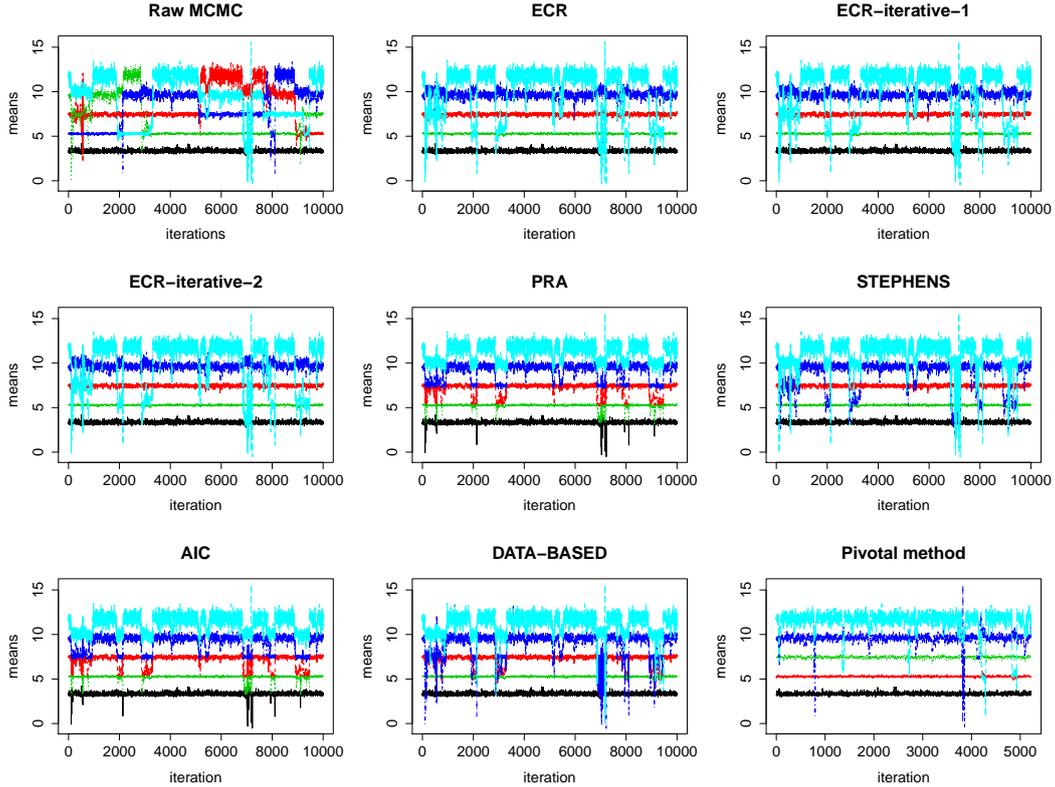}
\caption{Fishery data.  (\emph{From top left to bottom right}) Raw MCMC sample for $\mu_{g}, g =1,...,5$. Reordered MCMC samples by applying the permutations returned by \texttt{label.switching}
function, according to methods ECR, ECR-iterative-1, ECR-iterative-2, PRA, STEPHENS, AIC, DATA-BASED. Reordered MCMC samples according to the pivotal method.}
\label{8metodi}
\end{center}
\end{figure}

In Figure~\ref{8metodi} the raw MCMC sample (Top left) and the reordered MCMC samples for $\mu_{g}, g=1,...,5$, for different methods,  are shown. Despite an ordering constraint for the mean components (the priors are chosen according to the {\ttfamily independence} option, which favours a natural ordering of the means), label switching occurs, and the raw sampler is unable to yield useful means estimates for the single components. The \texttt{label.switching} function of the same package is used to reorder the obtained chains  according to the resulting permutations. The methods from the  \textbf{label.switching} package seem to perform similarly. In particular, for the greatest mean (light blue trace) there is a global tendency of switching. We note that for the DATA-BASED method the same happens also for the second mean (blue trace). Our pivotal method seems to work better in isolating the five high-posterior density regions. We recall that the reordering for our method is explained by \eqref{eq:relabelmu}.

Concerning the computational times reported in Table~\ref{tab:cpu}, AIC is the fastest method, since it only applies an ordering constraint and consequently permutes the simulated MCMC output, while STEPHENS ---a probabilistic relabelling--- is the slowest. Our method is quite fast, especially if compared with ECR-iterative-1, ECR-iterative-2 and DATA-BASED.

\begin{table}
\centering
\begin{tabular}{r|c}
Method & CPU time (sec.)\\
\hline
STEPHENS &  344.50 \\
PRA & 3.96  \\
ECR &  8.66 \\                              
ECR-iterative-1 &  60.83 \\
ECR-iterative-2 &  28.39 \\
AIC &  0.08 \\
DATA-BASED & 22.68 \\
\textbf{Pivotal} & \textbf{9.57}
\end{tabular}
\caption{Fishery data: CPU times in seconds for different methods, with $H=11000$, burn-in=1000, $n=256$ and $G=5$.}
\label{tab:cpu}
\end{table}
%
%


\section{Conclusions}
\label{conc}

We propose a simple procedure for dealing with label switching in Bayesian mixture models, based on the identification of as many pivots as mixtures components, used for relabelling the resulting MCMC chains. 
The main novelty of our contribution consists in providing some useful indications on how to choose the pivots, since, as mentioned in Section 6.1, the idea of solving the relabelling issue by fixing the groups for some units is not new \citep{chung2004difficulties}. We suggest to adopt one of six alternative methods based on a maximization or a minimization of some quantities derived from a similarity matrix
obtained through the MCMC sample, or a further demanding algorithm suitable when the number of groups $G$ is relatively  small (e.g. $G=4$). 

A fundamental issue is represented by the pairwise (perfect) separation between pivots, since it is crucial for the proposed 
procedure and, usually, non-trivial. 

From a computational side, the method appears to be fast and simpler than other relabelling methods, since it does not  require a maximization/minimization step at each iteration, and only requires a permutation of the labels induced by the pivots membership. 
A simulation study is conducted in order to test the proposed solution on different possible scenarios, showing overall good performances. A case study on a real dataset is presented, and the results seem to confirm the advantage of using the proposed methodology. Moreover, when also considering the computational time of our algorithm compared to some procedures available in the \textbf{label.switching} \texttt{R} package, we conclude that the proposed methodology may represent a
valid approach to the label switching problem and, in some cases, may be preferable to other existing solutions.

\bibliographystyle{chicago}

\newpage

\section*{Appendix}
\label{A:MUS}

\subsection*{MUS algorithm}

The algorithm of \textit{Maxima Units Search} is an alternative method for detecting pivots which does not rely upon a maximization/minimization step as the other six procedures in \eqref{eq:maxmeth} and \eqref{eq:minmeth}, but it searches for $\hat{G}$ pivots which satisfy a proper feature within the estimated similarity matrix $C$. The underlying idea is to choose as pivots those units in correspondence of which the $\hat{G} \times \hat{G}$ sub-matrix of $C$ with only the row and columns corresponding to $i_1,\ldots,i_{\hat{G}}$, is more often (close to) the identity matrix. Let us denote this sub-matrix of $C=(c_{ij})$ only containing the rows and columns corresponding to the pivots $i_1,\ldots,i_{\hat{G}}$ by $\mathcal{T}_{(\hat{G}\times \hat{G})}$. It is worth stressing that for a small number of groups (e.g., $G=4$) and a sample size $n$ ranging between $100$ and $1000$, this research can be computationally demanding. Furthermore, a positive number of identity matrices is not always guaranteed. However, the MUS algorithm has proved to be efficient in terms of mean square errors for group means estimation, as shown in Table~\ref{tab:MSE}. The main steps of the algorithm are summarized below.

\begin{description}
\item[(i)] For every group $g, \  g = 1,...,\hat G$, find  the \emph{maxima} units $j^{1}_{g},...,j^{M}_{g}$ within matrix $C$, i.e. the units in group $g$ with the greatest number of zeros in correspondence of the units of the other $\hat{G}-1$ groups, where $M$ is a precision parameter fixed in advance (in our simulation study $M=5$).
\item[(ii)] For these $M \times \hat G$ units, count the number of distinct identity sub-matrices of rank $\hat G$ $\mathcal{T}_{(\hat{G}\times \hat{G})}$ which contain them.
\item[(iii)] For each group $g, \  g = 1,...,\hat G$, select the unit which yields the maximum number of identity matrices of rank  $\hat G$. Such unit represents the pivot to be used for relabelling the chains as explained in Section~\ref{sec:pivotal}.
\end{description}

\end{document}